\documentclass[12pt,mathptm,times]{article}
\usepackage{graphicx}
\usepackage{setspace}
\usepackage[round,numbers]{natbib}
\usepackage{amsmath, amsthm, amsfonts, amssymb}
\usepackage{authblk}
\usepackage{amsmath}
\usepackage{amssymb}
\usepackage{graphicx}
\usepackage{mathrsfs}
\usepackage{enumerate}
\usepackage{pslatex}
\usepackage{bm}
\usepackage{color}
\usepackage{xcolor}
\usepackage[framemethod=TikZ]{mdframed}
\mdfdefinestyle{box}{%
    linecolor=black!60!white,
    outerlinewidth=1pt,
    roundcorner=10pt,
    innertopmargin=\baselineskip,
    innerbottommargin=\baselineskip,
    innerrightmargin=20pt,
    innerleftmargin=20pt,
    backgroundcolor=blue!10!white,
    font={\sffamily}}

\renewcommand{\ge}{\geqslant}
\renewcommand{\le}{\leqslant}

\newcommand{\toyLIFE}{{\fontfamily{qcr}\selectfont t\!\raisebox{-.1em}{o}\!y}%
{\fontfamily{phv}\selectfont \-LIFE}}
  
\title{On the networked architecture of genotype spaces and its critical effects on
molecular evolution}
\author[1,2]{Jacobo Aguirre}
\author[1,3]{Pablo Catal\'an}
\author[1,3,4,5]{Jos\'e A. Cuesta}
\author[1,2]{Susanna Manrubia}
\affil[1]{Grupo Interdisciplinar de Sistemas Complejos (GISC), Madrid}
\affil[2]{Programa de Biolog\'{\i}a de Sistemas, Centro Nacional de
Biotecnolog\'{\i}a (CSIC), Madrid, Spain}
\affil[3]{Dept. de Matem\'aticas, Universidad Carlos III de Madrid, Legan\'es,
Madrid, Spain}
\affil[4]{Instituto de Biocomputaci\'on y F\'\i sica de Sistemas Complejos
(BIFI)\\ Universidad de Zaragoza, Spain}
\affil[5]{UC3M-BS Institute of Financial Big Data (IFiBiD), Universidad Carlos
III de Madrid, Getafe, Madrid, Spain}

\date{\today}

\begin{document}


\maketitle


\begin{abstract}
Evolutionary dynamics is often viewed as a subtle process of change
accumulation that causes a divergence among organisms and their genomes.
However, this interpretation is an inheritance of a gradualistic view that has
been challenged at the macroevolutionary, ecological, and molecular level.
Actually, when the complex architecture of genotype spaces is taken into
account, the evolutionary dynamics of molecular populations becomes
intrinsically non-uniform, sharing deep qualitative and quantitative
similarities with slowly driven physical systems: non-linear responses
analogous to critical transitions, sudden state changes, or hysteresis, among
others. Furthermore, the phenotypic plasticity inherent to genotypes transforms
classical fitness landscapes into multiscapes where adaptation
in response to an environmental change may be very fast. The quantitative
nature of adaptive molecular processes is deeply dependent on a 
networks-of-networks multilayered structure of the map from genotype 
to function that we
begin to unveil. 
\end{abstract}

\section{Introduction}

Gradualism posits that any profound change in nature is the result of minor
cumulative modifications due to the action of slow but sustained processes.
First proposed in the framework of Geology at the end of the 18th Century by
James Hutton, gradualism underlies Charles Lyell's theory of
uniformitarianism~\cite{lyell:1830}, which formed one of the conceptual pillars
of Charles Darwin's evolutionary theory soon after~\cite{darwin:1859}. Ever
since, gradualism has been a powerful concept in the qualitative interpretation
of evolutionary change. 

The gradualistic view of evolution has been challenged at the macro- (fossil
record), meso- (ecological) and micro- (molecular) scales. In the 1970s,
analyses of data in the fossil record revealed an unanticipated pattern of
evolutionary stasis in the morphological change of species that was punctuated
by sudden jumps, leading to the theory of punctuated
equilibria~\cite{eldredge:1972}. The mechanistic models proposed to generate
that dynamical pattern are not unique, though the endogenous organisation of
the biosphere may have played a main role~\cite{sole:1996,sole:1997}. At present,
punctuated equilibrium is understood as an alternation of periods with
insignificant change (stasis) punctuated by rapid speciation, which may however
extend over a few hundred thousand years and result from complex evolutionary
dynamics~\cite{hunt:2012}.  Analogies between macroevolution and evolutionary
ecology were suggested on the basis that the degree of complexity observed in
the spatial and temporal organisation of both systems might be reflecting a
network-like organisation close to critical points~\cite{sole:1999}, the latter
resulting from a combination of external drivers and internal adaptive
responses. Research in this century has unveiled a large number of cases where
smooth environmental changes may indeed trigger sudden and irreversible
ecological responses~\cite{scheffer:2001}. The complex interaction between
natural systems and varying environments remains an open question of critical
relevance. The factors that make ecosystems respond smoothly or drastically to
a weakly evolving environment have attracted special interest, as there are
direct implications in the relation between humans and a changing biosphere
that could eventually reach a hazardous tipping
point~\cite{may:1977,scheffer:2001,barnosky:2012}.

\begin{figure}[h]
  \begin{center}
  \includegraphics[width=110mm]{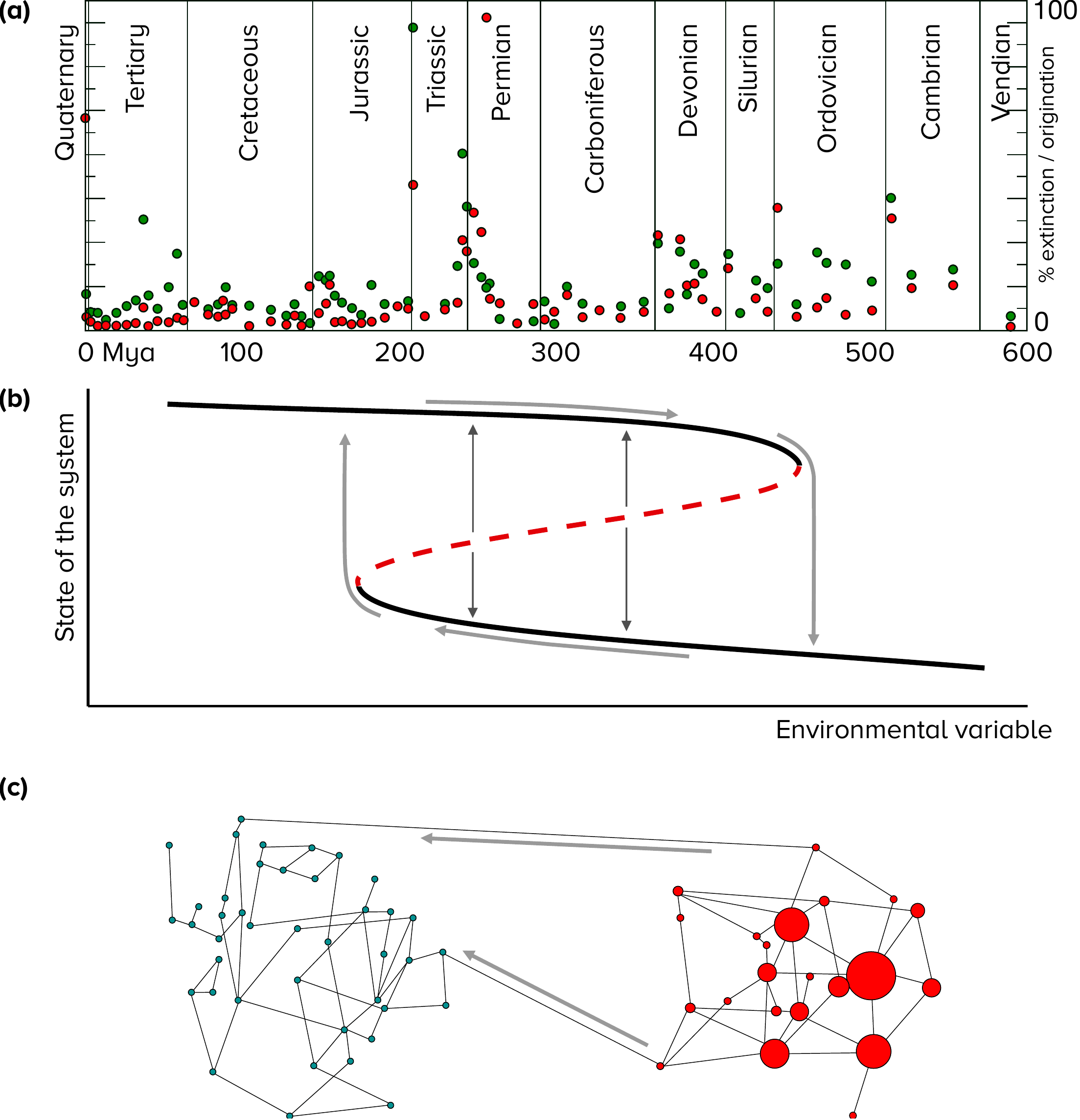}
\caption{\footnotesize Punctuated behaviour in macroevolution, ecology and molecular
dynamics. (a) Non-uniform pattern of extinctions (red symbols) and
originations (green symbols) in the last 610 Million years (0 is present). Each
point corresponds to a geological epoch, vertical lines separate geological
periods, as indicated. The vertical axis gives the percentage of
extinction/origination per estimated diversity at each epoch and per million
years. Data from~\cite{benton:1993}, geological epochs and
periods as in~\cite{harland:1990}. (b) Minor changes in environmental
variables might cause large, non-linear responses in the state of a variety of
systems. In some cases, two stable solutions (black curves) coexist with an
unstable solution (red curve) for a range of values of a control parameter. The
trajectories of systems might follow the path indicated by grey arrows as that
parameter increases, suffering a sudden jump from the upper to the lower
branch. Hysteretic behaviour appears and prevents the recovery of the initial
state when the environmental variable is reverted.  When the system is
initiated close to the unstable branch, it may attain any of the two possible
stable solutions (black thin arrows). (c) In the genotype space, nodes
represent genotypes and links correspond to single mutational moves.
Heterogeneous molecular populations contain a set of genotypes with variable
abundances, the latter represented through circle size. Fitter regions in
genotype space might be difficult to find if there are few mutational incoming
pathways (grey arrows). The population might be trapped in the red phenotype
for a relatively long time (stasis) as compared to the transition to the new
state once suitable mutations have appeared (punctuation).}
\end{center}
\label{fig:qnets}
\end{figure}

The formal description of non-uniform dynamics in natural systems is advancing
concomitantly with the number of examples supporting and clarifying the
theoretical framework (see figure~\ref{fig:qnets}). Shifts in ecosystems have been
formally described as bifurcations leading to hysteretic behaviour and also as
critical transitions. Analogous to fluctuations close to critical points, the
so-called early warning signals can anticipate such catastrophic
responses~\cite{scheffer:2009}. Empirical evidence of this phenomenon with a
single species has been described in laboratory populations of
yeast~\cite{dai:2012}, while there is a variety of well-documented examples in
ecology, such as the hysteretic loss and recovery of charophyte vegetation at
lake Veluwe~\cite{meijer:2000}, the desertification of the
Sahara~\cite{kassas:1995}, the loss of transparency in shallow
lakes~\cite{scheffer:1993} or the dynamics of woodlands in
Tanzania~\cite{dublin:1990}. A thorough description of this phenomenology is a
hard task, as it involves a wide variety of time scales and biological
levels ---many of them organised as complex networks--- that interact in a
complex manner~\cite{levin:1992}. At the molecular level, the architecture of
the genotype-phenotype map entails non-uniform evolutionary
dynamics~\cite{stadler:2001}. In particular, it has been shown that the steady
accumulation of point mutations under a selective pressure acting on the
phenotype yields population dynamics characterised by stasis (when sequences
explore neutral regions) punctuated by phenotypic changes (when a fitter
phenotype is found)~\cite{huynen:1996}.  Smooth changes at the level of
sequences do not preclude sudden adaptive changes at the level of function:
well-motivated models support that, like the state of ecosystems, changes in
genomic composition might be sudden, irreversible, and
unavoidable~\cite{aguirre:2015}. These dynamics have been also documented in
the {\it in vivo} evolution of a virus, influenza A, which shows a seasonal
pattern where expansion of genotypic diversity predates the finding and
fixation of strains with novel antigenic properties that escape immune
detection~\cite{koelle:2006,wolf:2006}.

Despite mounting evidence, the long tradition of relating small changes in
sequences to gradual changes in organisms and populations persists, often in a
tacit way. A significant example is Wright's adaptive
landscape~\cite{wright:1931}, which appears as a direct consequence of
gradualistic thought and counts amongst the most powerful metaphors in Biology,
one that has conditioned evolutionary thinking for almost a
century~\cite{svensson:2012}. Indeed, the image of a relatively smooth
landscape, where populations adapt by going uphill, are trapped in mountain
peaks and remain isolated from other possibly higher fitness maxima by deep
valleys, often appears as the way in which adaptation proceeds. This picture
implies a smooth and continuous genotype-to-phenotype (GP) map and a space of
low dimensionality. Thanks to advances in our knowledge of the molecular
structure of populations, we now know of important elements missing in most
theoretical adaptive landscapes. For example, genotypes of similar fitness are
found to form extensive networks that occasionally traverse the genotype space,
especially in spaces of high dimensionality~\cite{wagner:2011}. The GP map
actually entails a many-to-many correspondence: genotypes are plastic and may
yield different phenotypes when expressed in different environments. This
latter case seems to be much more common than previously thought, meaning that
the co-option of promiscuous, secondary gene functions~\cite{conant:2008} is
likely a common adaptive mechanism. From a formal viewpoint, therefore, the
complexity of the GP map implies that fitness landscapes should be visualised
as high-dimensional and interwoven sets of networks that unfold into multiple
layers under environmental change~\cite{catalan:2017}. New techniques, in
particular the use of deep sequencing and powerful massive ways to evaluate the
fitness of individual genotypes, represent a breakthrough in the empirical
characterisation of the complex genotype-to-phenotype-to-function
relationship~\cite{hinkley:2011,payne:2014b}. Interestingly, the
network-of-networks structure of genotype spaces described in realistic, though
artificial, models is also emerging in empirical characterisations of the
diversity of molecular populations~\cite{steinberg:2016}.

Adaptive evolutionary systems, such as large-scale evolution, ecology or
(molecular) populations, share deep analogies that can be likely ascribed to
their networked architecture plus a non-trivial relationship between exogenous
drivers and endogenous responses. In this review we will focus on molecular
dynamics, which is the least studied of those three profoundly entangled levels
of description of the evolutionary process. The architecture of genotype spaces
and the dynamics of evolving molecular populations are two sides of the same
coin. The heterogeneous structure of genotype spaces and its apparently
hierarchical organisation as a multilayer of networks of networks explains, among
others, punctuated dynamics~\cite{huynen:1996}, drift and switch
transitions~\cite{koelle:2006}, genomic shifts~\cite{aguirre:2015} or
Waddington's genetic assimilation~\cite{waddington:1953,catalan:2017}. 

\section{Genotype networks}
\label{sec:GN}

Kimura introduced the concept of neutral evolution in order to explain why many
mutations observed in RNA, DNA or proteins do not affect
fitness~\cite{kimura:1968,kimura:1984}. Neutrality implies that the GP map is
not one-to-one, but many-to-one, consistently explaining the high level of
polymorphism observed in natural populations. Soon after Kimura's seminal work,
navigability was hypothesised as an essential requirement to guarantee the
evolvability of molecular populations~\cite{maynard-smith:1970}. Usually,
navigability is believed to rely on the existence of sufficiently large {\it
neutral networks} (NNs) of genotypes~\cite{schuster:1994} since these should
permit the neutral drift of populations and a sustained exploration of
alternative phenotypes without a detrimental decrease in fitness. A NN is
formed by all genotypes that map into the same phenotype. As fitness is
linked to phenotype, all genotypes in a NN are implicitly assumed to have the
same fitness. Genotypes are the nodes of such networks, and links correspond to
single mutational moves. In its simplest and most popular definition, a
mutational move stands for a point mutation. Neutral networks can have one
or several connected components. Navigability on NNs has been subsequently
identified as a robust property of computational
models~\cite{huynen:1996,bastolla:2003,ciliberti:2007,matias-rodrigues:2011}
and natural molecular
populations~\cite{wolf:2006,schultes:2000,bloom:2007b,aguilar:2017}.

The actual set of genotypes visited by an evolving population, however, is
rarely neutral. Nearly-neutral mutations are common in finite
populations~\cite{ohta:1973}, augmenting their adaptive ability. In fact, any
finite mutation rate entails that populations are heterogeneous in sequence,
phenotype and function, such that the potential set of genotypes of a
population includes genotypes of different fitness, which constitute the actual
navigable network. In certain cases, as for ensembles of fast mutating
replicators such as quasispecies~\cite{eigen:1971,domingo:2006}, the
maintainance of a large phenotypic diversity and the permanent exploration of
the genome space become critical survival strategies~\cite{woo:2012}. We will
call {\it genotype network} the network of visited genotypes and, by extension,
any potentially navigable network in the space of genomes, regardless of the
fitness or phenotype of its nodes.

\subsection{Neutral networks in computational genotype-phenotype maps}

\begin{figure}[h]  
  \begin{center}
  \includegraphics[width=90mm,angle=270]{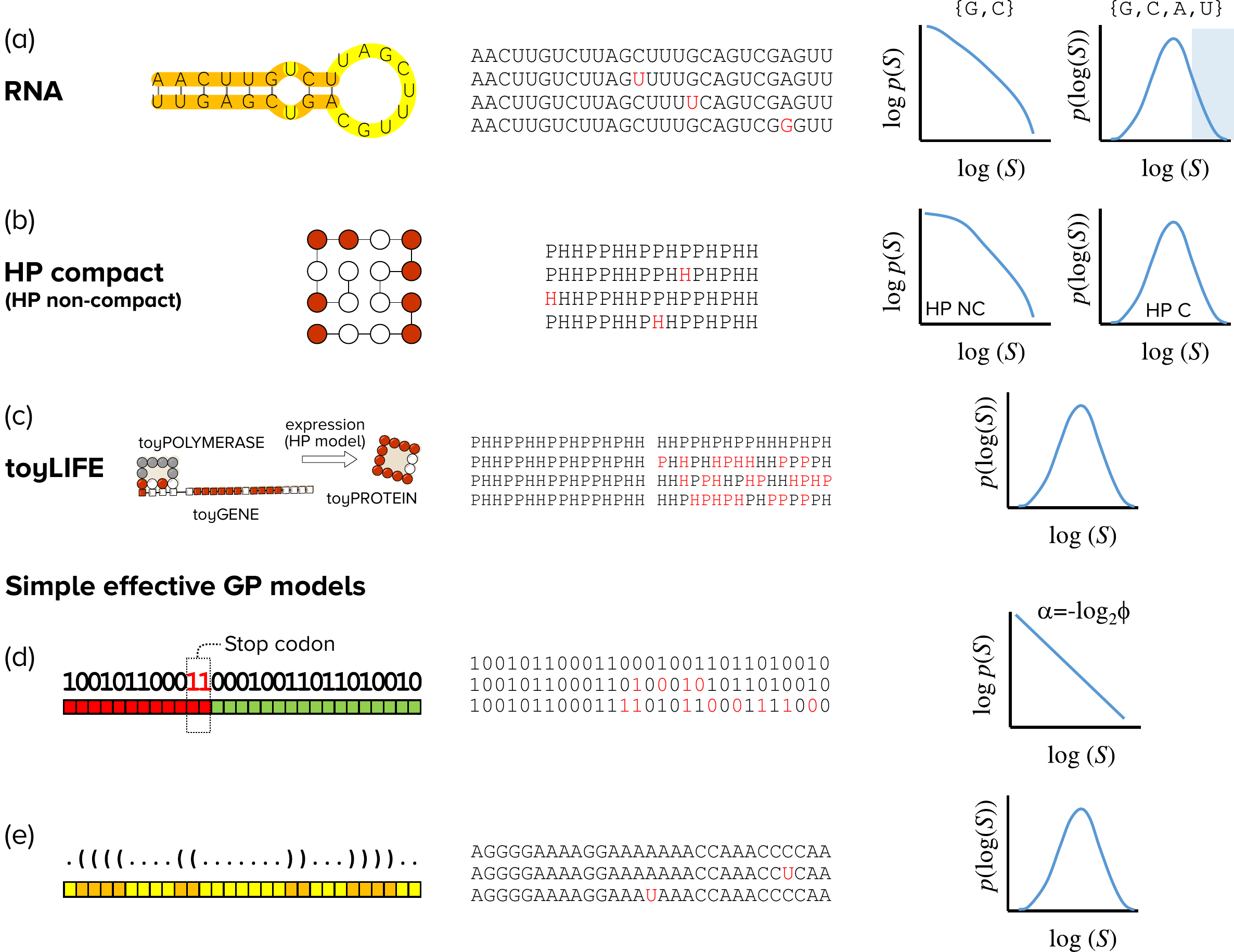}
\caption{\footnotesize Some examples of simple GP maps. For each model, and from left to right,
  we depict an example phenotype, some of the sequences in its neutral network
  (mutations that do not change the phenotype are highlighted in red), and the
  schematic functional form of the probability distribution $p(S)$ of phenotypes
  sizes $S$ found in computational or analytical studies. (a) RNA
  sequence-to-minimum-free-energy secondary structure. Mutations that do not
  disrupt the secondary structure appear with different probability in loops or
  stacks. In two-letter alphabets, the distribution of phenotype sizes is compatible
  with a power-law function~\cite{ferrada:2012} while, in four-letter alphabets, $p(S)$
  is well fit by a lognormal distribution~\cite{dingle:2015}. For long sequences,
  only the right-most part of $p(S)$ can be seen under random sampling of the genotype
  space~\cite{dingle:2015} (shaded). (b) The HP model, in its compact (as in the
  figure) or non-compact versions, has been studied as a model for protein folding. In
  non-compact versions, the distribution $p(S)$ has a maximum at $S=1$ and decays with
  a fat tail~\cite{irback:2002}, while in compact versions $p(S)$ resembles a lognormal
  distribution~\cite{garcia-martin:2018}. (c) \toyLIFE{} is a minimal model with several
  levels. HP-like sequences are read and translated to proteins that interact through
  analogous rules to break metabolites. The $p(S)$ of \toyLIFE{} is compatible with a
  lognormal distribution~\cite{arias:2014}. (d-e) Effective models where phenotype is
  defined in relation to the composition of sequences allow to analytically calculate the
  functional form of $p(S)$. Two examples are (d) Fibonacci's model~\cite{greenbury:2015},
  where $p(S)$ follows a power-law distribution and (e) an RNA-inspired
  model~\cite{manrubia:2017} which yields a lognormal distribution of $p(S)$.}
\end{center}
\label{fig:GPmodels}
\end{figure}

Neutral networks have been quantitatively characterised in a number of
computational GP maps (see figure~\ref{fig:GPmodels}). RNA sequences fold into a
minimum free energy secondary structure that we can take as a proxy for its
phenotype~\cite{hofacker:1994,schuster:1994}. Given a sequence length, the
number of minimum free energy secondary structures is much smaller than the
number of sequences, leading to large
NNs~\cite{schuster:1994,gruner:1996,fontana:1998,cowperthwaite:2008,jorg:2008,aguirre:2011,dingle:2015}.
In models of protein structure, such as the HP model~\cite{lau:1989}, proteins
are formed by strings of two amino acids: hydrophobic (H) and polar (P). As in
RNA, this sequence will fold into a minimum free energy structure, and there are
many more sequences than
structures~\cite{lipman:1991,li:1996,bornberg-bauer:1997,irback:2002}. In a
completely different model, gene regulatory networks possess an evolvable
architecture~\cite{crombach:2017} that gives rise to several
temporal gene expression patterns, which represent the phenotype. Again, many
interaction topologies representing the genotype give rise to a much smaller
number of gene expression patterns~\cite{wagner:2011,payne:2014a}. Neutral
networks also appear in metabolic processes. If we consider the genotype as a
list of enzymatic reactions and the phenotype as the set of metabolic sources
on which an organism can survive, it is found that many genotypes can actually
survive in a set of
environments~\cite{matias-rodrigues:2009,matias-rodrigues:2011,barve:2013,hosseini:2015}.
Finally, NNs have also been observed in complex models that include cellular
population dynamics and several levels from genotype to phenotype~\cite{ibanez:2014},
in more abstract GP maps, such as the polyomino model of polymer
self-assembly~\cite{johnston:2011,greenbury:2014}, \toyLIFE{} ---a multilevel
model of a simplified cellular biology~\cite{arias:2014,catalan:2018}---, and
in simplified combinatorial models~\cite{greenbury:2015,manrubia:2017}. 

Most NNs studied in the literature share a remarkable number of structural
properties~\cite{wagner:2011,ahnert:2017}:
\begin{enumerate}
\item Most phenotypes are rare, and only a few of them are very common.
Specifically, the probability of finding a phenotype when sampling uniformly at
random among all of them follows a lognormal distribution for a wide variety
of models~\cite{dingle:2015,catalan:2018,manrubia:2017} and a power law for
some special cases~\cite{ferrada:2012,greenbury:2015,manrubia:2017}. Therefore,
a small fraction of the largest phenotypes contains most genotypes, such that
in practice those are the only ones visible to natural
selection~\cite{cowperthwaite:2008,khatri:2009,dingle:2015}; together with the asymmetry in
the mutual accessibility of two phenotypes~\cite{fontana:1998,fontana:1998b}, that
property causes a form of (entropic) trapping in genotype
space~\cite{khatri:2009,schaper:2014,manrubia:2015,catalan:2018}.
\item The degree of a node in a NN, defined as the number of
one-mutant neighbours that belong to the same NN (aka its genotypic
robustness), is a heterogeneous quantity, although its distribution is often
unimodal~\cite{bastolla:2003,aguirre:2011,wagner:2011}. Additionally, the
average degree of a NN is proportional to the logarithm of the
size of the network~\cite{jorg:2008,aguirre:2011,greenbury:2016,manrubia:2017}.
\item These NNs are assortative, at least for phenotypes defined through
minimum-energy principles~\cite{bornberg-bauer:1999,wuchty:1999,aguirre:2011}.
In an assortative network, genotypes are connected to other genotypes of similar
degree, and this correlation in genotypic robustness causes
canalisation~\cite{ancel:2000}, leads to phenotypic
entrapment~\cite{manrubia:2015} and enhances evolvability~\cite{greenbury:2016}.
\item Neutral networks of common phenotypes percolate genotype space. In other
words, we can find two genotypes expressing the same phenotype with a sequence
similarity comparable to that of two randomly chosen
genotypes~\cite{gruner:1996b,wagner:2011}.
\item Most large phenotypes are one mutation away from each other, such that
genotypes yielding every common phenotype can be found at the boundary of any
large NN~\cite{gruner:1996b,bornberg-bauer:1997,wagner:2011,arias:2014}. As a
result, the search for new phenotypes among common ones is a fast process.
\end{enumerate}
The space of genotypes can be depicted in this context by a number of
interconnecting NNs when each node is projected in a horizontal (quasi-)
neutral layer whose vertical position represents its fitness value. In this
multilayer perspective~\cite{kivela:2014,boccaletti:2014}, intralayer
connections between individual nodes represent neutral mutations, while
interlayer connections represent mutations that beneficially (upwards) or
deleteriously (downwards) affect fitness~\cite{manrubia:2012}. It is however
important to keep in mind that this representation is suitable only if the GP
map is approximated as a many-to-one relationship, since it fails to include
the frequent correspondence between one genotype and several possible
(environment dependent) phenotypes, as will be discussed in
Sections~\ref{sec:promiscuity} and \ref{sec:promiscuity_theory}.

\subsection{Genotype networks in genotype-to-function maps}
The GP map is at best a toy representation of the relationship between genotype
and function, though it hopefully captures some of its statistical
properties. Computational studies suggest that structural properties of GP maps
are largely independent of the precise definition of
phenotype~\cite{li:2002,stich:2010} and of details of specific
models~\cite{wagner:2011,ahnert:2017}, and data to assess whether GP maps are a
sufficiently accurate representation of genotype-to-function maps ---which
represent a qualitative step forward--- is mounting. Advances in experimental
techniques have allowed to study the structure of the genotype-to-fitness
mapping through either experimental evolution
studies~\cite{poelwijk:2007,jimenez:2013,devisser:2014,devos:2015,steinberg:2016}
or high-throughput data~\cite{firnberg:2014,payne:2014b,aguilar:2017}. The
resulting experimental fitness landscapes confirm and extend the picture of
molecular evolution gained through the computational study of simple GP maps,
showing the presence of many quasi-neutral (eventually navigable)
regions~\cite{lauring:2011} and decaying correlations between phenotypes as the
mutational distance increases~\cite{kouyos:2012}. Natural fitness landscapes
have an intermediate degree of ruggedness, they are neither smooth nor random,
therefore revealing an important role of epistasis in shaping the topological
properties of genotype networks and in defining eventually accessible genomic
pathways for molecular
adaptation~\cite{poelwijk:2007,bank:2016,zagorski:2016,aguilar:2017}. 

Fitness landscapes have been theoretically explored through models where
phenotypes need not be explicitly defined and, instead, a fitness value is
associated to each genotype. This representation is closer to data retrieved
through empirical evolutionary experiments. The NK model~\cite{kauffman:1987} has
proved to be especially useful to generate an underlying landscape with
realistic degrees of ruggedness~\cite{koelle:2006,devisser:2014}. Furthermore,
it is relatively simple, only depending on two parameters ---the length of the
sequence $N$ and the level of ruggedness $K$--- but versatile enough to model
fitness landscapes with natural properties such as epistasis, multiple fitness
peaks and local optima~\cite{ostman:2014}. 

It turns out that topological differences between genotype networks, obtained
through data that map genotype to function and NNs, as described in the previous
subsection, are only cosmetic. It can be shown that spaces of genotypes endowed
with the structure of the NK model are also organised as a network of networks,
that is, as a set of genotype networks qualitatively equivalent to NNs connected
through a limited number of pathways~\cite{yubero:2017}. The structural properties
of genotype networks, visualised as a multilayer of networks of networks, define a
particular class of dynamics for populations evolving on such architecture.

The following sections are devoted to the not yet fully understood interaction
between the topology of genotype networks and the evolutionary dynamics of
heterogeneous populations ---at least from the formal viewpoint of dynamical
systems. We begin by synthesising current evidence to demonstrate that three
different dynamical situations (competitive transitions between different regions
of a NN~\cite{wilke:2001}, punctuated molecular adaptation~\cite{huynen:1996}, and
genomic shifts under varying environments~\cite{aguirre:2015}) can be described
within a unique conceptual and theoretical framework. In subsequent sections, we
will show how the latter framework can be extended to include the many-to-many
inherent structure of GP maps and environmental changes.

\section{Population dynamics on neutral networks}
\label{sec:dynamicsNN}

In order to describe mathematically the evolution of heterogeneous populations
on NNs, let us recall that many dynamical processes occurring on a
network of $m$ nodes can be expressed as

\begin{equation}
  \vec{n}(t)={\bf M} \vec{n}(t-1) = {\bf M}^t \vec{n}(0) \, ,
\label{DynDecomp}
\end{equation}
where $\vec{n}(t)$ is a vector whose components are the population of
individuals at each node at time $t$ and {\bf M} is an evolution matrix that
contains the particulars of the dynamical process (see~\textsf{BOX 1}).

\begin{mdframed}[style=box]
\begin{center}
BOX 1 -- {\bf Dynamics of replicators on a fitness landscape}
\end{center}
\vspace*{5mm}\noindent
The evolution of a population of asexually replicating individuals on a fitness
landscape described as a genotype network can be written as 

\begin{equation}
  \vec{n}(t)={\bf M} \vec{n}(t-1) = {\bf M}^t \vec{n}(0) = \sum_{i=1}^m
\lambda_i^t  (\vec{n} (0) \cdot \vec{u}_i) \vec{u}_i \, ,
\end{equation}
$\vec{u}_i$ and $\lambda_i$ are the eigenvectors and eigenvalues of the
evolution matrix {\bf M} and $m$ is the number of nodes of the genotype
network; $\vec{n}(t)$ has length $m$. We order the eigenvalues and eigenvectors
such that $\lambda_i\ge \lambda_{i+1}$. If {\bf M} is primitive,
Perron-Frobenius theorem for nonnegative matrices ensures that, over time, the
system evolves towards an asymptotic state characterised by the (unique) first
eigenvector $\vec{u}_1$. More precisely

\begin{equation}
\lim_{t \to \infty} (\lambda_1^t \alpha_1)^{-1} \vec{n}(t)=\vec{u}_1,
\qquad \alpha_1 = \vec{n} (0) \cdot \vec{u}_1>0,
\end{equation}
regardless of the initial condition $\vec{n}(0)$. The components of $\vec{u}_1$
(all of them guaranteed to be strictly positive by the same theorem) are
proportional to the fractions of the total population at each node once the
process has reached mutation-selection equilibrium, while its associated
eigenvalue $\lambda_1$ represents the asymptotic growth rate of the population.
The transient dynamics towards equilibrium is ruled by the subsequent
eigenvalues, but in most cases the time to reach the equilibrium state verifies
$t_{\text{eq}}\propto [\ln (\lambda_1/\lambda_2)]^{-1}$, since the
contributions of higher-order terms are suppressed exponentially
fast~\cite{aguirre:2009}. 

In a population of replicators that mutate with probability $0 < \mu < 1$ per
genotype and replication cycle, matrix {\bf M} can be decomposed
as~\cite{comment1}

\begin{equation}
  {\bf M} = (1 - \mu) {\bf F} + \frac{\mu}{S} {\bf G F} \, ,
\label{eq:primitive}
\end{equation}
where ${\bf F}$ is the diagonal matrix $F_{ij} = f_i \delta_{ij}$, $f_i$ being
the fitness (i.e., replication rate) of node $i$; ${\bf G}$ is the adjacency matrix
of a connected graph, whose elements are $G_{ij} = 1$ if nodes $i$ and $j$ are
connected and $G_{ij} = 0$ otherwise; and $S$ stands for the maximum number of
neighbours of a genotype~\cite{aguirre:2015}. When replicators are sequences of
length $l$ whose elements are taken from an alphabet of $A$ letters, the size
of the genotype space is $m=A^l$ and $S=l(A-1)$.

Matrices such as ${\bf M}$ in \eqref{eq:primitive} are guaranteed to be
primitive if the network ${\bf G}$ is connected and the diagonal of ${\bf F}$
is strictly positive.

Dynamics on a single NN is a particular case for which the fitness components
are $f_i = f$ if $i$ is a genotype in the NN and 0 otherwise ---all sequences
replicate at a rate $f$.
\end{mdframed}

For the sake of illustration let us start by considering a simple fitness
landscape with a single viable phenotype. The genotypes yielding the latter
constitute a NN and all remaining genotypes have zero fitness. Consider
genotypes as sequences of length $l$ whose elements are taken from an alphabet
of $A$ letters. Nodes represent different sequences and links connect those
sequences differing only in one letter. The evolution of a population
through the space of genotypes due to mutations is here limited to the NN ---or
to its largest connected component in case the NN is disconnected. An evolution
matrix that models such a dynamical process is

\begin{equation}
 {\bf M} = f(1 - \mu) {\bf I} + \frac{f\mu}{(A-1)l} {\bf G},
\label{Eq:M}
\end{equation}
where ${\bf I}$ is the identity matrix and  ${\bf G}$ is the adjacency matrix
of the connected network, with elements $G_{ij} = 1$ if nodes $i$ and $j$ are
connected, and $G_{ij} = 0$ otherwise. The genotypic robustness of a node is
proportional to its degree $k_i$, defined as the number of genotypes
one-mutation away that are on the network, $k_i = \sum_j G_{ij}$. ${\bf M}$
describes a population that every time step replicates at each node at a rate
$f>1$, each daughter sequence leaving the node with probability
$0 < \mu < 1$ and surviving with probability
$k_i\mu/(A-1)l$~\cite{aguirre:2009}, with $k_i$ the degree of the parental
node. If we define $k_{\text{min}}$, $k_{\text{max}}$, and $\langle k \rangle$
as the smallest, largest, and average degree of that NN respectively, we obtain
$k_{\text{min}}<\langle k \rangle \le \gamma_1 < k_{\text{max}}$ for any
heterogeneous network, $\gamma_1$ being the largest eigenvalue of the adjacency
matrix $\mathbf{G}$ \cite{boccaletti:2006}. In the case of two-letter alphabets,
$A=2$, $\gamma_1$ is bounded by the logarithm of the number of genotypes in a
NN~\cite{reeves:2016}. $\gamma_1$ also equals the average
degree of the population at equilibrium, $\kappa$, so the former inequality
implies $\kappa>\langle k \rangle$, indicating that the population selects
regions with connectivity above average on the NN. This fact shows a natural
evolution towards mutational robustness, because the most connected nodes are
those with the lowest probability of
experiencing lethal mutations. Nonetheless, the population might get trapped
in regions of lower connectivity if $N \mu<1$~\cite{nimwegen:1999}. The
tendency towards robustness does not preclude evolutionary innovation though.
On the contrary, NNs relevant in evolution spread on large regions in genome
space \cite{dingle:2015}, with the result that they can be more robust and at
the same time more evolvable~\cite{wagner:2008,draghi:2010,greenbury:2016}. A
positive correlation between neutrality and evolvability stems from the the
fact that NNs are very interwoven: for example, all common RNA
structures of length $l$ can be found within a small radius of a randomly
chosen sequence in genotype space ---a property known as ``shape space
covering''~\cite{gruner:1996b,reidys:1997}. The mutual proximity of NNs in
genome space (the so-called NN
apposition~\cite{fontana:1998,fontana:2002}) has been observed empirically. Two
remarkable examples are ribozymes and viruses. Indeed, two RNA sequences with
independent origins can fold and function as different ribozymes when their
sequences are forced to evolve to increase their similarity, eventually
differing in only two nucleotides~\cite{schultes:2000}; diffusion on NNs
is instrumental to permit innovation and immune escape in
influenza~A~\cite{koelle:2006}.

The eigenvectors of the adjacency matrix ${\bf G}$ are also eigenvectors of the
evolution matrix ${\bf M}$, as can be seen in Eq.~\eqref{Eq:M}. Their
respective eigenvalues, $\gamma_i$ and $\lambda_i$, are different ---albeit
related through $\lambda_i=f(1-\mu)+\gamma_if\mu/(A-1)l$. As a consequence,
in NNs the asymptotic state of the system only depends on the topology of the
NN, and parameters such as the mutation rate $\mu$ or the sequence
length $l$ exclusively affect the transient dynamics towards
equilibrium~\cite{nimwegen:1999,aguirre:2009}.  This result cannot be
extrapolated to more general fitness landscapes, where both the equilibrium
state of the population and the transient dynamics depend in a non-trivial
fashion on network topology and genotype fitness~\cite{aguirre:2009} (c.f.
Eqs.~(2) and~(4) in \textsf{BOX 1}). 

Heterogeneity in the degree of the nodes, or equivalently in genotypic
robustness, and the assortativity inherent to many NNs have
important consequences in the dynamics of populations. Soon after the
hypothesis of the molecular clock~\cite{zuckerkandl:1965} was put forward,
variations in genotypic robustness were suggested as an explanation for its
unexpected overdispersion~\cite{takahata:1987}. If networks are furthermore
assortative, the probability that the population leaves the network diminishes
the longer the time spent on it, leading to a progressive (phenotypic)
entrapment.  Beyond a systematic increase in the overdispersion of the process
with time, assortativity entails an acceleration in the fixation rate of
neutral mutations~\cite{manrubia:2015}, invalidating the Poissonian assumption
underlying the molecular clock.

\section{Punctuated dynamics in molecular adaptation}
\label{sec:punctuated}

As soon as more realistic architectures of the genotype space are considered,
dynamics becomes punctuated. This fact has been highlighted in formal studies
stating that GP maps based on RNA sequence-to-structure relationship
naturally imply punctuation, irreversibility and modularity in phenotype
evolution~\cite{stadler:2001}, and has been nicely illustrated in computational
works~\cite{huynen:1996,fontana:1998,fontana:2002}. 

The formal scenario that we use here starts at the level of genotypes, but also
takes into account the non-trivial topology induced by the mapping onto
phenotypes. By means of techniques that exploit the networked and modular
structure of genotype spaces, we will show that the dynamical behaviour is
qualitatively similar in three different situations, that is if (i) a NN has
two or more regions of high connectivity linked through few possible mutational
pathways, (ii) a population encounters a phenotype of fitness higher than the
extant one, or (iii) mutation-selection equilibrium is perturbed through an
environmental change that entails a modification of the
fitness landscape. Underneath the punctuated dynamics observed in those
situations there is a common mechanism: a (formal) competition between regions
with a high internal connectivity that are sparsely connected to one another.
These highly internally connected regions may be different clusters of genotypes in a
single NN, different phenotypes each characterised by its own NN, or
different regions in a fitness landscape. Actually, this synthesis emerges as a
generalisation of processes occurring on a wide variety of biological,
technological and social dynamics on networks of networks (i.e. networks
connected through a limited number of connector links). This class of processes
admits a description in terms of competitive scenarios where each network is
defined as an independent agent struggling with the rest for a particular kind
of resource~\cite{aguirre:2013,buldu:2016,iranzo:2016}: {\it eigenvector
  centrality} (see~\textsf{BOX 2}).

\begin{mdframed}[style=box]
\begin{center}
BOX 2 -- {\bf When networks of networks compete for centrality}
\end{center}
\vspace*{5mm}

\noindent
In complex network theory, the eigenvector centrality $x_k$ of a node $k$
in a network is defined as the $k$th component of the eigenvector of its
adjacency matrix $\mathbf{G}$ corresponding to the largest eigenvalue
$\gamma_1$~\cite{newman:2010}. The eigenvector centrality has become the most
extended metric for node importance because of its wide range of applications,
which include Google Pagerank~\cite{langville:2006}, estimations of the
professional impact of scientists~\cite{senanayake:2015} and
journals~\cite{bergstrom:2007}, the importance of individuals in a social
group~\cite{seary:2003} or of regions in the brain~\cite{lohmann:2010}, and
dynamical processes such as disease or rumour spreading (see~\cite{newman:2010}
for an overview).

This measure can be generalised to other dynamical processes if ${\bf
G}$ is replaced by another (nonnegative) matrix ${\bf M}$: the new eigenvector
centrality is defined through $\vec{u}_1$, the eigenvector corresponding to
$\lambda_1$, the largest eigenvalue of $\mathbf{M}$ (see e.g.~\textsf{BOX 1}).
In evolutionary dynamics, the eigenvector centrality is thus the fraction of
population with each genotype at mutation-selection
equilibrium~\cite{aguirre:2009}. We use this generalisation in the following.

When several interconnected networks compete for centrality, the winnings of
each competing network $\alpha$ are calculated as the total centrality
$C_{\alpha}$ accumulated by all its nodes

\begin{equation*}
C_{\alpha}=\sum\limits_{j \in\alpha} u_{1,j}/\sum\limits_{k=1}^m u_{1,k} \, , 
\end{equation*}
where $j$ runs on the nodes of network $\alpha$ and $m=\sum_{\mu} m_{\mu}$ is
the total number of nodes in the network of networks. The outcome of such
confrontations for centrality and the time needed by the winner to prevail
drastically depend on (i) the internal structure of the competing networks
$\alpha=1,\dots,K$, as characterised by their maximum eigenvalue
$\lambda_{1,\alpha}$, in a way that networks with larger $\lambda_{1,\alpha}$ in
general obtain more centrality than their competitors, and (ii) the connector
nodes, that is, the boundary nodes that connect one of these networks with the
rest of them through connector links.

When connector links occur only through nodes with little centrality
(aka peripheral connections), almost all centrality remains in the network
with the largest eigenvalue $\lambda_{1,\alpha}$. If for some reason (e.g.~an
environmental change) the eigenvalue of a different network overcomes
$\lambda_{1,\alpha}$, a sharp centrality redistribution takes place. The time to
reach the equilibrium significantly increases close to that transition.
\end{mdframed}

\subsection{Metastable states and punctuation in a network-of-networks
architecture}

In Section~\ref{sec:dynamicsNN} we have focused on the dynamics of populations
evolving on a single NN characterised by a well-defined region of maximum
connectivity. Under those conditions,
the evolutionary dynamics of a sufficiently large population is smoothly
canalised towards the maximally connected region of the
NN~\cite{nimwegen:1999,ancel:2000,aguirre:2009} ---something that has
measurable effects on the fixation rate of neutral
mutations~\cite{manrubia:2015}.  However, there is no {\it a priori} reason to
assume that generic NNs do not present a complex structure formed by more than
one cluster of nodes with high internal connectivity and sparse connections to
one another. If this is so, the evolutionary dynamics of populations on NNs can
display an alternance of metastable states (which might appear as true
equilibria at short times) with periods where neutral mutations are rapidly
fixed~\cite{wilke:2001}. 

\begin{figure}[h]
\centering
  \includegraphics[width=90mm]{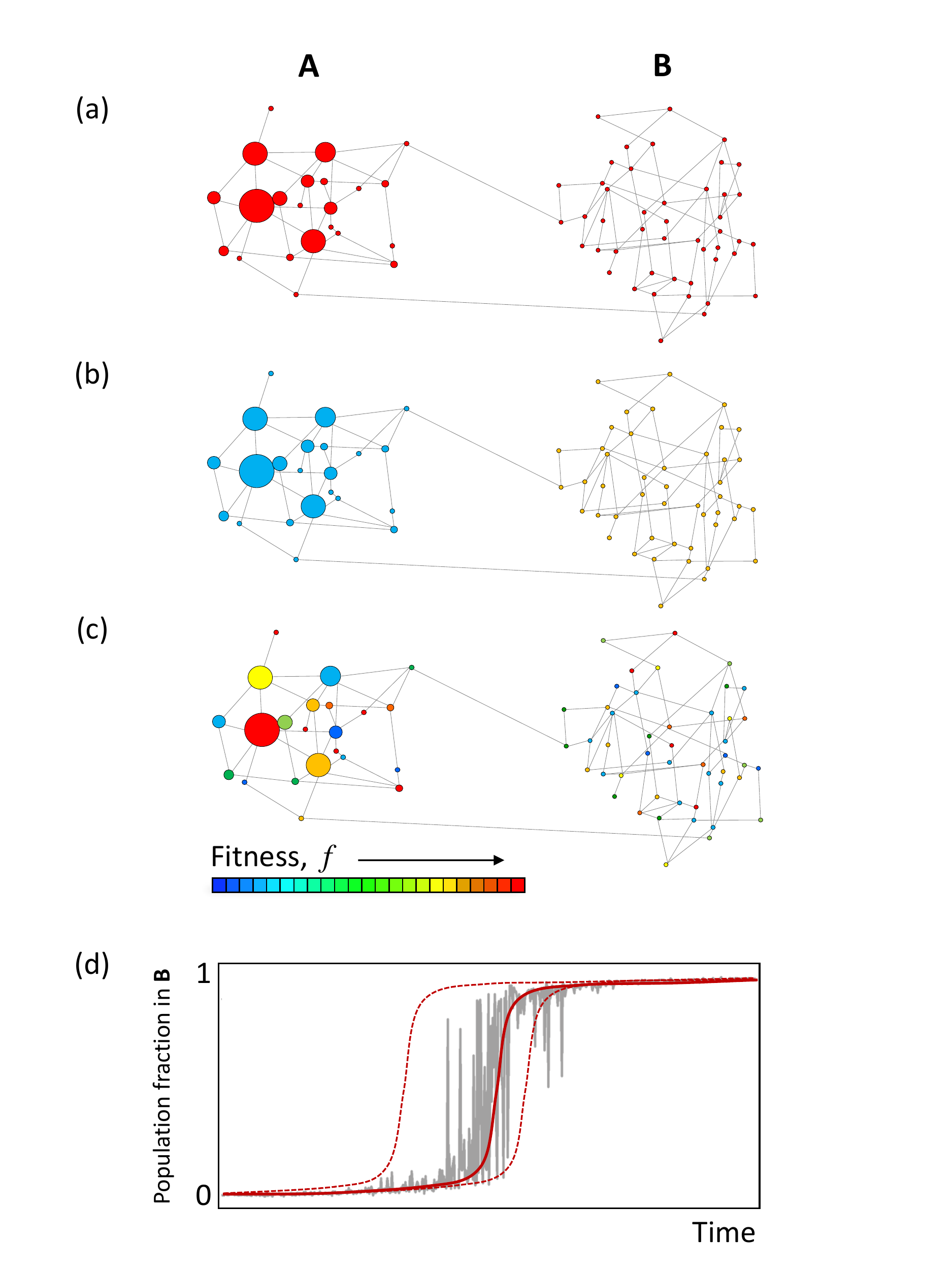}
\caption{\footnotesize Genomic shifts result from the network-of-networks structure of the space
  of genotypes. 
  Without loss of generality, we assume that $\lambda_{1,A} < \lambda_{1,B}$
 and the whole population is initially in network {\bf A}. In
  (a-c), colours indicate the fitness of each node, as shown by the colour
  scale, and circle size is indicative of the number of individuals at
  each node.
  Though nodes in network {\bf B} are represented with small circles, we assume
they have no population initially. (a) Two weakly-coupled regions of a unique
NN. Differences in their
  eigenvalues only depend on differences in their topology. (b) Two different NNs with
  different fitness. The effect of fitness and topology can be separated, both affect
  their eigenvalues. (c) Two weakly-connected regions in a fitness landscape. The effects
  of fitness and topology cannot be decoupled. 
  (d) In all cases, the time of transitions is a stochastic variable, but the
  transition is fast once the mutational pathway is found (red curves, corresponding
to different realisations of the process). In changing
  or noisy environments, the fitness value of each sequence might vary in time, so
  transitions are decorated by fluctuations (grey curve) whose strength grows as the
  tipping point is approached.}
\label{fig:NN-communities}
\end{figure}

The formalism that describes competition between networks for centrality, while
originally introduced in the framework of complex network theory, was recently
proven to be fully applicable to the study of populations evolving in the space
of genotypes~\cite{yubero:2017}. The population distribution at
mutation-selection equilibrium is given by the first eigenvector $\vec{u}_1$ of
the matrix ${\bf M}$ that characterises the dynamical process, and therefore
the centrality that each network competes for coincides with the fraction of
organisms that populate its corresponding sequences in the asymptotic state.
In general, the most populated network in the equilibrium is the one 
with the largest eigenvalue $\lambda_1$ of matrix ${\bf M}$
(\textsf{BOX 2}).

Let us illustrate in the simplest case how a population moves from a subnetwork
with a lower eigenvalue $\lambda_{1,A}$ to a subnetwork with a larger eigenvalue
$\lambda_{1,B}$ in the framework of competition for centrality.
Figure~\ref{fig:NN-communities}a represents two regions of a NN weakly connected.
As previously described, we have $\lambda_{1,A}=f(1-\mu) +\gamma_A f \mu/(A-1)l$,
and similarly for network {\bf B}. Note that the latter network will be
eventually attracting the population if the eigenvalue corresponding to its
evolution matrix $\lambda_{1,B}$ is larger than that of {\bf A},
and as a consequence if the same applies for the adjacency matrices (i.e.
$\gamma_B > \gamma_A$). This result shows that the separating barrier only
depends on the topological structure
(size and connectivity) of each subnetwork. The transition to a region with higher
connectivity occurs upon stochastic appearance of mutations along connecting
pathways. This process is highly contingent, so the time of the punctuation is
difficult to predict (red lines in figure~\ref{fig:NN-communities}(d)). Actually,
too small populations might be indefinitely trapped in regions as
{\bf A}~\cite{huynen:1996}. 

\subsection{Drift and switch dynamics in adaptive transients}

Early evidences of punctuation in molecular adaptation came from computational
simulations of populations of RNA sequences evolving towards a target secondary
structure~\cite{huynen:1996}. Typically, populations remain on the current
phenotype until a higher-fitness solution is found, that is, until one of the
genotypes in the population acquires a mutation that produces a new, fitter
phenotype. This event is preceded by a ``search'' in the original phenotype
during which the population accumulates neutral mutations and increases its
genotypic diversity. The switch transition is not deterministic, since
different phenotypes can be reached first depending on the stochastic
occurrence of mutations. Once the new phenotype has been found, the transition
occurs exponentially fast but, concomitantly, the population experiences a severe
bottleneck that reduces its genotypic diversity. In this scenario, a new
phenotype can be accessed through any genotype in the neighbourhood of
genotypes of the original phenotype, though peripheral genotypes (those with a
higher number of links pointing to different phenotypes, i.e.~of low
robustness) are more likely to act as connectors than highly robust, central
genotypes~\cite{manrubia:2015}. This drift and switch dynamics is
characteristic of any realistic GP map with a structure such as that described
in Section~\ref{sec:GN}. In the dynamical framework of competition between
networks, each phenotype represents now a distinguishable network characterised
by its size, connectivity and fitness level. Connector links correspond to
regions of apposition between the two networks, which exist in most cases (in
particular when the two phenotypes considered are common) but are difficult to
find if populations are finite due to the vastness of genotype spaces and
NN~\cite{catalan:2017}.
Also, the connector links might join regions with similar fitness but different
internal connectivity, or regions with different fitness, among many other
possibilities. Different paths to adaptive improvement are taken with different
probability. For example, narrow neutral paths are crossed much faster than
fitness valleys~\cite{nimwegen:2000}.

Figure~\ref{fig:NN-communities}b illustrates the situation of two phenotypes
with different fitness values (i.e. replicative ability of its nodes) coupled
through narrow paths. The transition to phenotype {\bf B} might occur if 
$\lambda_{B,1}>\lambda_{A,1}$ which implies that

\begin{equation}
\frac{f_A}{f_B}<\frac{1-\mu+\mu\gamma_B/(A-1)l}{1-\mu+\mu\gamma_A/(A-1)l}
\approx 1+\frac{\mu}{(A-1)l}(\gamma_B-\gamma_A), \quad \mu\ll 1,
\end{equation}
where the specific effect of fitness $f_i$ and topology $\gamma_i$ is quantified.

The survival-of-the-flattest effect represents one
particular case of such competition where the two competing regions have different
levels of fitness, different mutation rates (a situation that can be easily included
in the framework above), and different levels of
robustness~\cite{wilke:2001b,codoner:2006}, which effectively accounts for
different topologies~\cite{wilke:2003}.
Epochal evolution (i.e. metastable states punctuated by rapid transitions to
fitter states) have also been observed in evolutionary search algorithms, as
referred to a class of optimisation
techniques~\cite{nimwegen:2000b,nimwegen:2001}.

The theory can be easily extended to any number of phenotypes in competition and
yields a clear prediction regarding the phenotype that will be eventually
attracting the population. The largest eigenvalue of any matrix {\bf M},
$\lambda_1$, is a fundamental quantity that synthesises information on the
topology of the underlying network, on the fitness of its nodes, and on the
mutation rate. These three elements combine in a non-trivial way to determine the
competitive ability of a population on a given network. In this respect, a
population can asymptotically displace a competitor for a number of different
reasons, namely because (i) it spreads on a larger NN, (ii) its average fitness is
higher, (iii) it spreads on a network with higher connectivity, (iv) it mutates at
an advantageous rate with respect to its competitors, or (v) any suitable
combination of the previous reasons. 

\subsection{Smooth environmental changes and genomic shifts}

There is empirical evidence that environmental changes affect the evolutionary
dynamics of populations and their eventual fate~\cite{fuentes:2015}. Recalling that
fitness is an environment-dependent quantity, environmental changes can be formally
cast as modifications of the fitness associated to genotypes. When a genotype space
is mapped to a realistic fitness landscape, smooth environmental changes can be
represented as gradual modifications of the fitness value of each genotype.
Since phenotype is here a hidden variable, at this point we do not need to
consider possible changes in phenotypic expression due to environmental variation.
This possibility will be discussed later though.

Even if environmental variations are smooth, populations may eventually suffer
sudden transitions in their genomic composition~\cite{aguirre:2015}. In the
case of finite populations, there is a non-zero probability of extinction if the
pathway linking the (decreasingly fit) current state of the population to a new
region populated by fitter phenotypes is not found sufficiently
fast~\cite{yubero:2017}. The abundance and breadth of connecting pathways
depends on the roughness of the landscape and on the fraction of lethal
mutations, which can be put in correspondence with important variables such as
the degree heterogeneity of the corresponding genotype networks and the
holeyness of the landscape~\cite{gavrilets:2004}. These quantities tune the
number of connector links between different regions with significant fitness and the
centrality of their connector nodes. As a consequence of the above, fitness
landscapes can be described as a network of networks formally analogous to the
examples discussed previously (see figure~\ref{fig:NN-communities}c).

Early warning signals that forecast the proximity of tipping points (and
therefore of a putative extinction threshold) can be defined in analogy to
studies of sudden shifts in ecology~\cite{scheffer:2009}. Close to those state
transitions populations show flickering and hysteresis, i.e.~a dependence
on its previous states that causes trapping and metastability, and is
eventually responsible for extinction~\cite{yubero:2017}.

Summarising, facing evolutionary systems from the viewpoint of competing
networks turns the space of genotypes into a network of networks at several
different levels. The full consequences of this architecture are still to
be understood, though they are certainly far from trivial: relevant phenomena
such as robustness~\cite{buldyrev:2010,gao:2011},
synchronisation~\cite{um:2011,aguirre:2014},
cooperation~\cite{gomezgardenes:2012,wang:2012,iranzo:2016}, or epidemic
spreading~\cite{dickison:2012,saumell:2012,yagan:2012} exhibit different
features when their dynamics occur on a single network or on a network of
networks. 

\section{The many-to-many nature of the GP map}
\label{sec:promiscuity}

Our discussion so far has assumed that each genotype corresponds to a unique
phenotype. Adaptation to a new environment or selection pressure, therefore,
has to be achieved through mutations, and we have discussed some of the
non-trivial phenomena that appear when heterogeneous populations evolve in a
complex genotype space. However, there are many cases in which genotypes
express more than one phenotype, opening up new possibilities for adaptation:
in any realistic realisation, the GP map is many-to-many, since genotypes are
able to express different phenotypes in a variety of situations.   
In this section we present several examples of this phenomenon and discuss how
it alters the dynamics discussed in previous sections. The reader should know
that the level of formal description achieved is poorer than for dynamics on
networks and has received much less attention up to now. Our feeling is that,
as shown in previous sections, theory should help towards unifying processes
and concepts that are treated at present as different phenomena. However, the
following sections rely much more on the description of the latter than on
quantitative results. A full mathematical formalism that describes at once the
multilayered, network-of-networks structure of genotype-to-function map is
an open and on-going problem of the highest relevance. 

\subsection{Molecular promiscuity}

Enzymes were classically thought to be highly specific:
one enzyme--one substrate--one reaction. However, recent experimental data has
shown that, in fact, many enzymes are able to catalyse more than one reaction,
a phenomenon that has been termed catalytic or functional
promiscuity~\cite{obrien:1999,copley:2003,babtie:2010,khersonsky:2010,copley:2015,copley:2017}.
This means one amino acid sequence corresponds to more than one phenotype.
Promiscuous enzymes are not hard to find in sequence space. For example, single-site
mutants of bacterial enolases can actually perform secondary functions not found in
the wild type, while maintaining their original activity~\cite{schmidt:2003}.
Moreover, these promiscuous functions are easily evolvable: enzymes can accumulate
mutations that do not alter their main function, but which change radically their
secondary ones~\cite{aharoni:2005,amitai:2007,bloom:2007}, and the activity of
secondary functions can be increased several orders of magnitude with very few
mutations~\cite{obrien:1999,yang:2016,baier:2017}.

Promiscuous activities can help enzymes evolve toward new functions. A polymorphic
population of enzymes can diversify with respect to its secondary functions if they
bear no fitness costs to the organism, leading to the accumulation of what has been
termed cryptic genetic variation~\cite{paaby:2014}. When selection 
pressure for a new function appears, those enzymes in the population that carry
out that function as a promiscuous activity will be already functional and, in
a sense, pre-adapted for it. The new function can then be improved through
over-expression \cite{khersonsky:2010} or gene duplication that liberates one
copy of the enzyme to specialise in the new function \cite{hughes:1994,
obrien:1999, copley:2015}. These promiscuous activities also have an effect on
metabolism, connecting different metabolic pathways \cite{kim:2010,
copley:2015}, and therefore enabling their gradual evolution: promiscuous
enzymes can develop their secondary functions, so that certain steps in a
pathway become more efficient, in turn liberating other enzymes to focus on
other parts of the pathway. The evolution of metabolic pathways, therefore, can
be achieved in a more parsimonious way. When a new pathway is needed, cells with
promiscuous enzymes will maybe perform the needed reactions, and give these
sequences an adaptive advantage.

Functional promiscuity is not restricted to enzymes: transcription factors have
been shown to bind many different motifs with comparable binding energies
\cite{payne:2014b, copley:2015, aguilar:2017}.  Also, proteins can be
mistranslated \cite{bratulic:2015}, a process that is several orders of
magnitude more common than genetic mutations, and thus at a given moment in
time, some proteins will have a different amino acid sequence, with potentially
different functions that can accelerate adaptation to a new function
\cite{whitehead:2008, drummond:2009, yanagida:2015}. Some protein sequences
will be more likely to yield new functions under these phenotypic mutations. 

Promiscuity is also not restricted to proteins. Early computational work on RNA
secondary structures \cite{schuster:1994} already suggested that RNA molecules
could fold into more than one structure, and recent experimental studies have
found evidence of RNA molecules that can perform more than one different
function \cite{vaidya:2009, talini:2011}. The best examples are ribozymes (RNA
enzymes) that are able to catalyse two different reactions \cite{schultes:2000,
lau:2009, hayden:2011}. Computational \cite{ancel:2000, wagner:2014} and
experimental studies \cite{hayden:2011} suggest that secondary functions in RNA
molecules can evolve as easily as in proteins, and that this functional
promiscuity can spread through populations as cryptic genetic variation,
accelerating the rate at which new functions are found in evolution. Even if
these functions are performed marginally at first, they will give the sequence
an advantage if they are selected for, and freedom to improve the new
function in genotype space. In fact, theoretical models predict that promiscuous
functions can help accelerate evolution towards a new function, through what
has been called the look-ahead effect \cite{whitehead:2008}. Although this
phenomenon was originally proposed for phenotypic mutations, it is also valid
for promiscuous enzymes and RNA molecules.

\subsection{Phenotypic heterogeneity and bet-hedging}

The fact that one sequence can perform more than one function is not restricted
to the molecular level. At the regulatory level, for instance, expression noise
is very common \cite{elowitz:2002, raser:2005, maheshri:2007}, due to the
stochastic nature of transcription and translation and the small number of
molecules involved in these processes. Expression noise leads to phenotypic
heterogeneity \cite{ackermann:2015, vanboxtel:2017}, where two genetically
identical genotypes can, under the same conditions, express two different
phenotypes at the cellular level. Although expression noise is inherent to the
biochemical process of building the phenotype from the genotype, cells can
control it to some level \cite{raser:2005, herranz:2010, lehner:2008,
little:2013}, and they can also use it to their advantage \cite{eldar:2010,
ackermann:2015, vanboxtel:2017}. For instance, genotypes can evolve a
stochastic switching mechanism that enables them to alternate between two
different phenotypes, a phenomenon that has been termed bet-hedging
\cite{veening:2008}. At a given moment in time, a fraction of the population
will express one phenotype and the rest another one. Each phenotype is typically
advantageous in one environment and disadvantageous in another, and so the
ability to switch between them is adaptive under some conditions
\cite{norman:2015}. Typical examples of bet-hedging are bacterial competence
\cite{suel:2006} and persistence \cite{balaban:2004}. Bet-hedging is a common
mechanism that can also emerge in evolution experiments~\cite{beaumont:2009}.
These strategies would not be possible without functional promiscuity.

\subsection{Phenotypic plasticity}

Another piece of this puzzle comes from phenotypic plasticity, a well-known
phenomenon in which a genotype is able to express different phenotypes in
different environments~\cite{west-eberhard:2003}. Notice the difference from
phenotypic heterogeneity as discussed above: phenotypic plasticity is only
unveiled when an environmental cue appears. In fact, strategies such as
bet-hedging arise when the cost of developing a plastic response ---which is
able to sense the environment--- is so high that it becomes
disadvantageous~\cite{norman:2015}.

Phenotypic plasticity has been known for a long time in multicellular
organisms, but it appears at the unicellular and molecular level as well.
Proteins are not only promiscuous: they can also carry out different functions
in different environments, a phenomenon that is called moonlighting
\cite{jeffery:1999, copley:2012}. One classical example are crystallin lenses,
enzymatic proteins whose function becomes structural when expressed at very
high concentrations \cite{piatigorsky:2007}. The same gene can also express
different proteins through alternative splicing \cite{copley:2012}. RNA
molecules can fold into different structures at different temperatures,
performing different functions \cite{kortmann:2012}. RNA thermometers, as they
are called, can be designed computationally \cite{garcia-martin:2016}. Gene
regulatory networks have different spatio-temporal expression patterns when
exposed to different environmental inputs~\cite{payne:2014c,
espinosa-soto:2011, jimenez:2017}, and metabolic systems are able to survive on
different food sources~\cite{matias-rodrigues:2009, barve:2013, hosseini:2015}.

A plastic population will be able to automatically survive in a new
environment, if it expresses a viable phenotype. Once in the new environment,
it might spread through the new fitness landscape, maybe losing its
original plasticity. Many theoretical and computational studies of plasticity
and its relationship with adaptation have been proposed \cite{via:1985,
scheiner:1993, ancel:2000b, price:2003, lande:2009, scheiner:2012,
gomez-mestre:2013}, although most of them do not include the complexities of
the GP map that we have discussed in our previous sections.
They assume that phenotypes that are close in trait value to the ones present
in the population will always be achievable through mutations. Therefore, the
discussion of when and how phenotypic plasticity will be promoted cannot
account for the biases induced by more or less abundant phenotypes, asymmetric
connections between them and other factors discussed so far in this review,
which could affect how easily plasticity is developed. There are, however, some
computational studies that explicitly model GP maps, focusing
on RNA molecules \cite{ancel:2000} and gene regulatory networks
\cite{espinosa-soto:2011, draghi:2012}.

\section{Hints for a dynamical theory of many-to-many GP maps}
\label{sec:promiscuity_theory}

\subsection{Promiscuity redefines the fitness landscape}

How do we integrate all of this data into the framework we have been discussing
so far in this review? The presence of phenotypic noise or functional promiscuity
(at the molecular or regulatory level) implies that a single genotype, in a
given environment, will express more than one phenotype in a probabilistic manner.
Therefore, the effective fitness of the genotype will be an intermediate value
related to the fitness associated to each phenotype. Na\"ively, one could guess that
the fitness $f_i$ of sequence $i$ would be $f_i=\sum_{p \in \mathcal P} f(p)\pi_i(p)$,
where $\mathcal P$ is the set of all phenotypes, $f(p)$ is the fitness of phenotype
$p$, and $\pi_i(p)$ is the probability that sequence $i$ expresses phenotype $\pi$
(alternatively, $\pi_i(p)$ represents the fraction of the homogeneous population
with genotype $i$ expressing phenotype $\pi$). To illustrate one such case, consider
a population of RNA sequences that perform their function by interacting with a
ligand. Under the minimum free energy mapping usually considered in the literature,
all RNA sequences expressing the optimal structure as their minimum free energy
are assigned the same fitness. Including promiscuity, however, alters this fitness
function. Two sequences belonging to the same NN have different compositions, and
this variation leads, in general, to differences in their folding energies and also
in the repertoire of structures with which they are
compatible~\cite{mccaskill:1990}. Differences in the folding energy entail
differences in the average time spent in the minimum free energy secondary structure
for each specific sequence. In this situation, a more accurate definition of fitness
takes it as proportional to the time spent in the optimal secondary structure.
Therefore, two sequences belonging to the same NN have different fitness values under
this more realistic quantification of their function.

However, a careful investigation of the underlying (stochastic) population
dynamics reveals that the simple average above is not of general applicability, as
the next example illustrates. Consider a homogeneous population of cells expressing
a certain phenotype with probability $p$, and another one with probability $1-p$.
The replication rate $\beta$ of both phenotypes is the same, but the second 
phenotype has a higher death rate $\delta_2 > \delta_1$---i.e. it has a lower
fitness, defined as the difference between birth and death rates, $f=\beta-\delta$.
There is no mutation in this example. Whenever any cell replicates, the daughter
cell expresses one of the two phenotypes with the aforementioned probabilities,
regardless of the mother's phenotype. Calling $m_1(t)$ and $m_2(t)$
the number of cells of each type at time $t$, we can use results from birth-death
processes theory to derive the following system of ordinary differential equations:

\begin{equation}
  \begin{pmatrix}\dot{m}_1(t) \\ \dot{m}_2(t)\end{pmatrix} =
  \begin{pmatrix}\beta p - \delta_1 & \beta p \\ \beta (1-p) & \beta
    (1-p) - \delta_2 \end{pmatrix} \begin{pmatrix} m_1(t)
    \\ m_2(t) \end{pmatrix}.
\end{equation}
We diagonalize the system to obtain its largest eigenvalue
(and thus, the asymptotic fitness of the population):

\begin{equation}
  \lambda_1 =\frac{\beta-\delta_1-\delta_2}{2}+
    \sqrt{\frac{(\beta-(\delta_2-\delta_1))^2}{4}+\beta
      p(\delta_2-\delta_1)}.
  \label{eq:fitnessbd}
\end{equation}
With some algebra, we can show that $\lambda_1 > 
(\beta-\delta_1)p+(\beta-\delta_2)(1-p)$, the latter being the result of the
na\"{\i}ve guess above, i.e. that the average fitness of the population is the
weighted average of the fitness of the visited phenotypes, where weights are the
probability that a genotype expresses each phenotype. The discrepancy arises, in
this case, because cells expressing the second phenotype die more often. As a
result, the population has an overrepresentation of cells expressing the more
stable phenotype: their fraction in the population is actually greater than $p$. 

Despite the differences between the two examples discussed, it appears that the
effect of promiscuity can be accounted for by properly redefining the fitness
landscape. Each example, however, will need to be carefully examined to correctly
translate its dynamical details to a suitable definition of fitness. 

\subsection{Dynamics of plastic phenotypes under frequent environmental
changes}

Phenotypic plasticity means that the same genotype expresses different phenotypes
in different environments, such that different evolution matrices have to be
considered in each of the environments (see~\textsf{BOX 3}). To fix ideas, suppose
we have two different environments alternating every generation, with
associated matrices $\textbf M_1$ and $\textbf M_2$. Then the evolution of the
population will be given by the largest eigenvalue of the matrix $\textbf M_2
\textbf M_1$ and asymptotic state of the population turns out to be an orbit with
period 2, as long as some conditions are fulfilled. Both matrices (and their product)
must be primitive (see~\textsf{BOX 1}). This happens, for instance, if all
nodes have positive fitness or if, after removal of the zero-fitness nodes,
none of the two networks breaks down into different connected components.
If this condition is not met the asymptotic state will depend on the initial
condition. Likewise, even if all nodes have positive fitness but the fitness of some
of them is very small, the population can get trapped in metastable states for very
long times. But one can also imagine that alternating environments can have the
opposite effect, namely, that the transit of certain pathways strongly hindered in
both environments when kept constant may be facilitated by their alternation.

\begin{mdframed}[style=box]
    \begin{center}
BOX 3 -- {\bf Dynamics of replicators on a shifting fitness landscape}
\end{center}
\vspace*{5mm}\noindent
The framework introduced in \textsf{BOX 1} can be extended to account for
environmental changes. For the sake of simplicity we will just consider the case
in which the environment alternates between two states, but generalisations of
this are self-evident. The fitness of every node needs not be the same in each
environment, and as a result the evolution matrices of both
environments (we will denote them by $\mathbf M_1$ and $\mathbf M_2$) will be different.

Let us begin by exploring the case in which, starting in environment 1, we
alternate environments every generation. Then the equation for the
evolution of the population reads

\begin{equation}
  \vec{n}(t)=
\begin{cases}
[\mathbf M_2\mathbf M_1]^{t/2}\vec{n}(0), & \text{$t$ even,} \\
[\mathbf M_1\mathbf M_2]^{(t-1)/2}\mathbf{M}_1\vec{n}(0), & \text{$t$ odd.}
\end{cases}
\end{equation}
This means that, in general, the evolution of the population will be dominated
by the largest eigenvalue of the matrix $\mathbf M_2\mathbf M_1$ at even times
and of the matrix $\mathbf M_1\mathbf M_2$ at odd times, regardless of
$\vec{n}(0)$. (Starting from environment 2 would only swap the parity of times,
but not the general results.)

Interestingly, the eigenvalues of cyclic permutations of a product of matrices
are the same, and the corresponding eigenvectors are easily related to each
other. Thus, if $\lambda_1$ is the largest eigenvalue of $\mathbf M_2\mathbf M_1$
and $\vec{v}_1$ its corresponding eigenvector, then the eigenvector of matrix
$\mathbf M_1\mathbf M_2$ will be $\mathbf M_1\vec{v}_1$, so the asymptotic
population will grow as $\lambda_1^{t/2}$ and the fraction of population
will cycle through

\begin{equation}
\vec{v}_1 \quad\rightarrow\quad
\frac{\mathbf M_1\vec{v}_1}{|\mathbf M_1\vec{v_1}|} \quad\rightarrow\quad
\vec{v}_1.
\end{equation}

The case in which environments change following a random pattern is particularly
interesting. In this case
\begin{equation}
  \vec{n}(t)=\mathbf{M}^t\vec{n}(0), \qquad \mathbf M\equiv
\left\langle\prod_{k=1}^t\mathbf M_{\mu_k}\right\rangle^{1/t},
\end{equation}
where $\mu_k\in\{1,2\}$ is a discrete random process whose dynamics is
prescribed (for instance, it can take each of the two values with a certain
probability, or $\mu_1$ can take any value with a certain probability and swap
every time step with another probability). The expected value is to be taken
over realisations of this process. The largest eigenvalue of $\mathbf M$ and
its corresponding eigenvector will determine the asymptotic behaviour of the
population. Mathematically, this process is not fully characterised yet, but
it is not difficult to carry out its numerical implementation.
\end{mdframed}

\begin{figure}[h]
  \includegraphics[width=140mm]{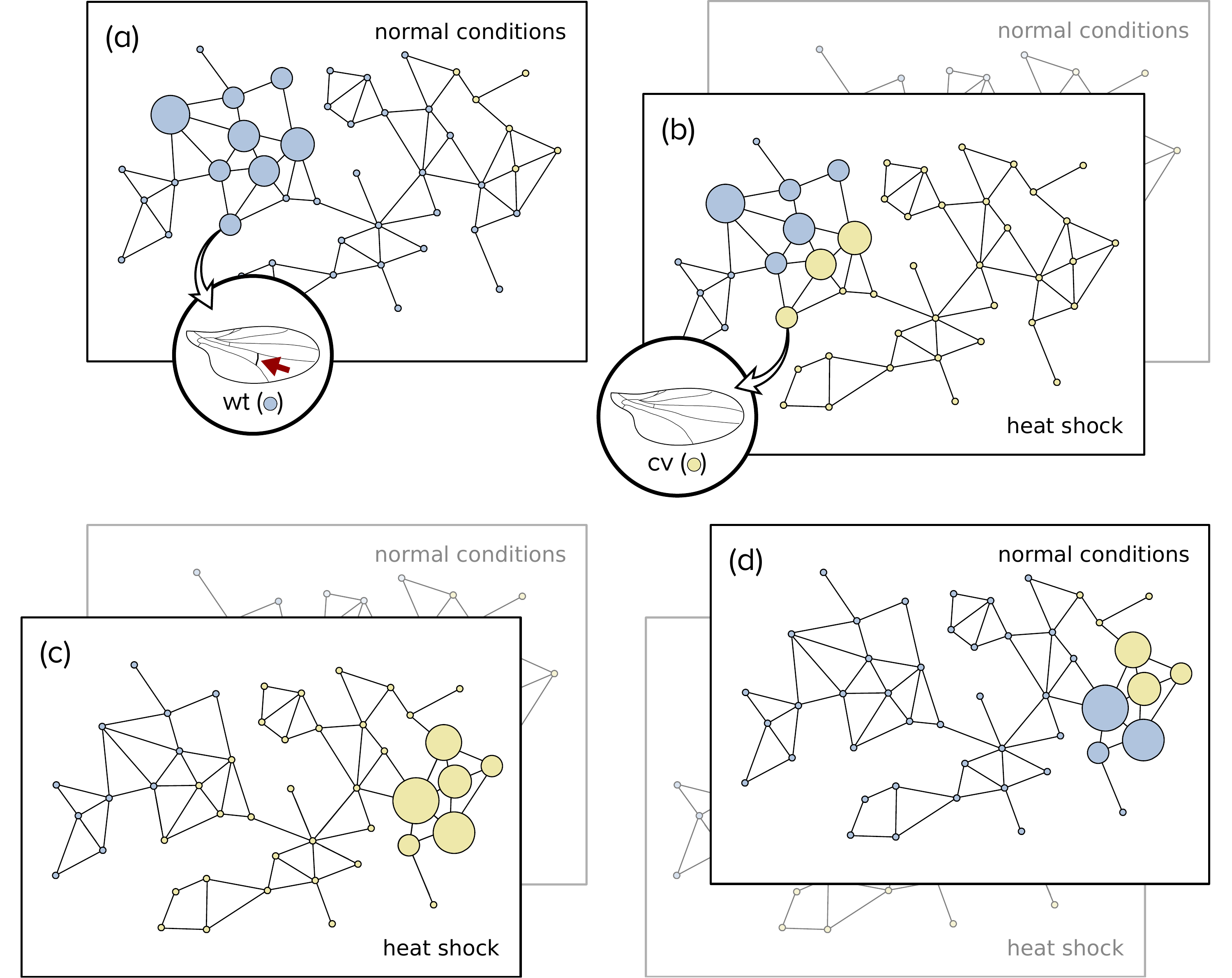}
\caption{\footnotesize Waddington's genetic assimilation under the light of genotype
  networks.
  Each layer of the network
  represents a different environment. Here there are two environments:
  normal conditions and heat shock. As in previous figures, circle
  size is proportional to the number of individuals populating that
  node ---small circles represent unpopulated nodes. The
  colour of each node represents now its phenotype, instead of its
  fitness. Note that every genotype appears in both layers, and that
  connections between them are the same in both environments: the only
  property that changes is the phenotype.
(a) A population of flies develops wings with a cross-vein (the
  wild-type phenotype, wt, blue) when bred in normal conditions. (b) When
  exposed to heat shock during development, some of the flies in the
  original population develop new wings without cross-veins (the
  cross-veinless phenotype, cv, yellow). (c) Breeding the flies under
  heat shock and then selecting for those flies expressing the
  cross-veinless phenotype, the population drifts towards a new part of
  genotype space, exploring a new neutral network (or possibly
  increasing fitness in the new environment). (d) After some time, the
  population is bred again in normal conditions, and some flies in the
  population keep expressing the cross-veinless phenotype. Their
  phenotype has been genetically assimilated.}
\label{fig:genshift}
\end{figure}

This analysis can be extended to more complicated alternating patterns of the
two environments, the only differences being that the asymptotic state will
exhibit a longer period. For instance, if environments change according to the
pattern 112112112\dots, and $\lambda_1$ and $\vec{v}_1$ are the largest eigenvalue
and its corresponding eigenvector of the matrix $\mathbf M_2 \mathbf M_1^2$,
then the population will grow as $\lambda_1^{t/3}$ and the fraction of population
will cycle through

\begin{equation*}
\vec{v}_1 \quad\rightarrow\quad
\frac{\mathbf M_1\vec{v}_1}{|\mathbf M_1\vec{v}_1|} \quad\rightarrow\quad
\frac{\mathbf M_1^2\vec{v}_1}{|\mathbf M_1^2\vec{v}_1|} \quad\rightarrow\quad
\vec{v}_1.
\end{equation*}

A qualitative representation of this idea was already proposed in the form of
adaptive multiscapes~\cite{catalan:2017} (see figure~\ref{fig:genshift}). It was 
shown there that the evolutionary phenomena introduced by phenotypic plasticity,
such as Waddington's genetic assimilation~\cite{waddington:1953}, could be
easily understood in terms of a multilayered network of genotype networks.
Genetic assimilation is a very interesting phenomenon. In Waddington's
experiment, a plastic population of flies was exposed to a new environment, in
which they expressed a different phenotype (called cross-veinless). They were
selected for this new phenotype under the new environment, so they spread
through the genotype network in the way we have discussed in
Section~\ref{sec:punctuated}. After some time, when the population was brought
back to the original environment, some of the individuals kept the
cross-veinless phenotype, instead of reverting to the wild-type
(figure~\ref{fig:genshift}). The phenotype that originally appeared only
plastically was now being expressed without environmental changes: it had 
become genetically assimilated. Adaptive multiscapes help in the qualitative
understanding of the molecular mechanisms underlying genetic assimilation, among
others, since 
the population dynamics sketched in \textsf{BOX 3} suffice to explain it.

\section{Discussion and prospects}

A large body of current evidence shows that the gradualistic view of evolution
is at odds with the mechanisms operating at the molecular level, where
discontinuous changes and fast pre-adaptations are the rule rather than the
exception. We have presented three basic mechanisms with a strong effect on the
evolutionary dynamics of biomolecules: fast exploration of new phenotypes by
heterogeneous populations spread over neutral networks, competition between
different networks for population (the evolutionary counterpart of eigenvalue
centrality) and plasticity of phenotypes. But ubiquitous and general as they
may be, these are by no means the only ones. Several other mechanisms and
phenomena have been left out from our framework.

The first one has to do with mutations. The most parsimonious change in a
genome is represented by point mutations. All through this review we have shown
how even these minor changes frequently cause major phenotypic modifications.
The evolution of genomes, however, is often driven by mutational mechanisms
that substantially modify them, such as gene duplication or horizontal gene
transfer (HGT). The latter will potentially cause effects of magnitude larger
than point mutations, and therefore entail still stronger effects on phenotypes
and functions. The structure of genomes, especially the existence of universal
regularities in the distribution of genomic elements~\cite{koonin:2011b} speaks
about dominant mechanisms beyond organismal adaptation~\cite{iranzo:2014,
iranzo:2017}. Gene sharing through HGT has played a main role in the adaptation
of microorganisms~\cite{schonknecht:2014} and is so common in microbial
evolution that it has led to the idea of network genomics~\cite{koonin:2015}.
The reconstruction of gene-sharing networks for viruses~\cite{iranzo:2016b} has
uncovered a hierarchical and modular structure that drastically changes our
view of viral species as well-defined entities.  Instead, the topology of such
networks reveals an utmost plastic system where genes behave as highly mobile
pieces, and where not only adaptation but also evolutionary innovations might
be strongly promoted through combinatorial processes ---especially in viruses
with segmented genomes~\cite{lucia-sanz:2017}. This plastic view of the genome
can be straight forwardly extended to cellular organisms.

Secondly, we have not included any kind of sexual reproduction nor
recombination ---of which HGT is a particular case. Though recombination might
slow-down evolution under strong selection~\cite{ueda:2017}, in most of its forms
it is a powerful enhancer of the search for novelty~\cite{peabody:2017}. This power
is very well illustrated in experiments of DNA shuffling \cite{crameri:1998}, where a
chimaeric cephalosporin created from recombination of four different ones
achieves a 270-fold increase of resistance to antibiotic ---compared to the
8-fold increase achieved by the best cephalosporin created through point
mutations alone. On top of that, the interplay between recombination and the
genotype-phenotype map may induces a fascinating disruptive dynamics that
resembles sympatric speciation \cite{azevedo:2006}, so speciation ---one of
evolution's major themes--- may not be properly understood unless recombination
is suitably incorporated in our dynamical models. However, this cannot be done
if size- and frequency-dependent evolution operators are not introduced,
because the probability that a recombination event takes place depends on the
relative presence in the population of the sequences to be recombined. The lack
of a suitable framework to describe this complication leaves any `ecological'
interaction between molecules or genes out of the picture. This is probably the
weakest point of the network formalism ---one that is of paramount importance
to tackle in future work.

Even if we constrain ourselves to the range of applications to which the
formalism we are advocating for does apply, its actual implementation is not
free from serious difficulties. To begin with, the vastness of genotype spaces
makes it impossible to explore any realistic genotype-phenotype map in depth.
This is a handicap that will not be solved with more powerful computers, so
we need to turn to an alternative description of evolutionary dynamics.
Fortunately, all models of the genotype-phenotype map share a set of common
properties regardless of the details. This situation is similar to the one
faced by Statistical Physics in its aim to go from microscopic models to
macroscopic description, and so it can be dealt with in a similar vein. If
details do not matter, we may try to build a mesoscopic description in which
phenotypes, rather than genotypes, are the basic elements of our dynamical
framework, and in which microscopic details are subsumed in an effective,
possibly non-Markovian stochastic dynamics~\cite{manrubia:2015}.

We also need to figure out how to incorporate promiscuity and environment in
our evolutionary picture, in a way that does not require to run specific
simulations for each particular case. If a mesoscopic description is to be
made, any change in the environment would entail a full reconfiguration of the
network of phenotypes, thus affecting not only the phenotype that the
population currently occupies, but also the transitions between different
phenotypes ---hence the evolutionary pathways. A way to incorporate the effect
of the environment would be through a multilayer formalism for networks
\cite{kivela:2014,boccaletti:2014}, where different layers would correspond to
different environments. Generalising the dynamics described here to a
multilayer network is as yet an open problem.

\section{Acknowledgements}

This work has been supported by the Spanish Ministerio de Econom\'{\i}a y
Competitividad and FEDER funds of the EU through grants ViralESS
(FIS2014-57686-P) and VARIANCE (FIS2015-64349-P). JA is supported through
grant SEV-2013-0347. PC is supported through the European Union's YEI funds.



\begin{thebibliography}{205}
\providecommand{\natexlab}[1]{#1}
\providecommand{\url}[1]{\texttt{#1}}
\expandafter\ifx\csname urlstyle\endcsname\relax
  \providecommand{\doi}[1]{doi: #1}\else
  \providecommand{\doi}{doi: \begingroup \urlstyle{rm}\Url}\fi

\bibitem[Lyell(1830)]{lyell:1830}
C.~Lyell.
\newblock \emph{Principles of geology, being an attempt to explain the former
  changes of the Earth's surface, by reference to causes now in operation}.
\newblock London: John Murray, 1830.

\bibitem[Darwin(1859)]{darwin:1859}
C.~Darwin.
\newblock \emph{On the Origin of Species by Means of Natural Selection, or the
  Preservation of Favoured Races in the Struggle for Life}.
\newblock John Murray, London, 1st edition, 1859.

\bibitem[Eldredge and Gould(1972)]{eldredge:1972}
N.~Eldredge and S.~J. Gould.
\newblock Punctuated equilibria: an alternative to phyletic gradualism.
\newblock In T.~J.~M. Schopf, editor, \emph{Models in Paleobiology}, pages
  82--115. San Francisco: Freeman Cooper, 1972.

\bibitem[Sol{\'e} and Manrubia(1996)]{sole:1996}
R.~V. Sol{\'e} and S.~C. Manrubia.
\newblock Extinction and self-organized criticality in a model of large-scale
  evolution.
\newblock \emph{Phys.\ Rev.\ E}, 54:\penalty0 R42, 1996.

\bibitem[Sol\'e et~al.(1997)Sol\'e, Manrubia, Benton, and Bak]{sole:1997}
R.~V. Sol\'e, S.~C. Manrubia, M.~J. Benton, and P.~Bak.
\newblock Self-similarity of extinction statistics in the fossil record.
\newblock \emph{Nature}, 388:\penalty0 764--767, 1997.

\bibitem[Hunt et~al.(2012)Hunt, Hopkins, and Lidgard]{hunt:2012}
G.~Hunt, M.~J. Hopkins, and S.~Lidgard.
\newblock Simple versus complex models of trait evolution and stasis as a
  response to environmental change.
\newblock \emph{Proc.\ Natl.\ Acad.\ Sci.\ USA}, 112:\penalty0 4885, 2012.

\bibitem[Sol\'e et~al.(1999)Sol\'e, Manrubia, Benton, Kauffman, and
  Bak]{sole:1999}
R.~V. Sol\'e, S.~C. Manrubia, M.~J. Benton, S.~Kauffman, and P.~Bak.
\newblock Criticality and scaling in evolutionary ecology.
\newblock \emph{Trends Ecol.\ Evol.}, 14:\penalty0 156--160, 1999.

\bibitem[Scheffer et~al.(2001)Scheffer, Carpenter, Foley, Folke, and
  Walker]{scheffer:2001}
M.~Scheffer, S.~Carpenter, J.~A. Foley, C.~Folke, and B.~Walker.
\newblock Catastrophic shifts in ecosystems.
\newblock \emph{Nature}, 413:\penalty0 591, 2001.

\bibitem[May(1977)]{may:1977}
R.~M. May.
\newblock Thresholds and breakpoints in ecosystems with a multiplicity of
  stable states.
\newblock \emph{Nature}, 269:\penalty0 471, 1977.

\bibitem[Barnosky et~al.(2012)Barnosky, Hadly, Bascompte, Berlow, Brown,
  Fortelius, Getz, Harte, Hastings, Marquet, Martinez, Mooers, Roopnarine,
  Vermeij, Williams, Gillespie, Kitzes, Marshall, Matzke, Mindell, Revilla, and
  Smith]{barnosky:2012}
A.~D. Barnosky, E.~A. Hadly, J.~Bascompte, E.~L. Berlow, J.~H. Brown,
  M.~Fortelius, W.~M. Getz, J.~Harte, A.~Hastings, P.~A. Marquet, N.~D.
  Martinez, A.~Mooers, P.~Roopnarine, G.~Vermeij, J.~W. Williams, R.~Gillespie,
  J.~Kitzes, C.~Marshall, N.~Matzke, D.~P. Mindell, E.~Revilla, and A.~B.
  Smith.
\newblock {Approaching a state shift in Earth's biosphere}.
\newblock \emph{Nature}, 486:\penalty0 52, 2012.

\bibitem[Scheffer et~al.(2009)Scheffer, Bascompte, Brock, Brovkin, Carpenter,
  Dakos, Held, van Nes, Rietkerk, and Sugihara]{scheffer:2009}
M.~Scheffer, J.~Bascompte, W.~A. Brock, V.~Brovkin, S.~R. Carpenter, V.~Dakos,
  H.~Held, E.~H. van Nes, M.~Rietkerk, and G.~Sugihara.
\newblock Early-warning signals for critical transitions.
\newblock \emph{Nature}, 461:\penalty0 53, 2009.

\bibitem[Dai et~al.(2012)Dai, Vorselen, Korolev, and Gore]{dai:2012}
L.~Dai, D.~Vorselen, K.~S. Korolev, and J.~Gore.
\newblock {Generic indicators for loss of resilience before a tipping point
  leading to population collapse}.
\newblock \emph{Science}, 336:\penalty0 1175--1177, 2012.

\bibitem[Meijer(2000)]{meijer:2000}
M.~Meijer.
\newblock \emph{Biomanipulation in the Netherlands: 15 years of experience}.
\newblock Wageningen Univ., Wageningen, The Netherlands, 2000.

\bibitem[Kassas(1995)]{kassas:1995}
M.~Kassas.
\newblock {Desertification: a general review}.
\newblock \emph{J. Arid Env.}, 30:\penalty0 115--128, 1995.

\bibitem[Scheffer et~al.(1993)Scheffer, Hosper, Meijer, Moss, and
  Jeppesen]{scheffer:1993}
M.~Scheffer, S.~Hosper, M.-L. Meijer, B.~Moss, and E.~Jeppesen.
\newblock {Alternative equilibria in shallow lakes}.
\newblock \emph{Trends Ecol.\ Evol.}, 8:\penalty0 275--279, 1993.

\bibitem[Dublin et~al.(1990)Dublin, Sinclair, and McGlade]{dublin:1990}
H.~T. Dublin, A.~Sinclair, and J.~McGlade.
\newblock Elephants and fire as causes of multiple stable states in the
  {Serengeti-Mara} woodlands.
\newblock \emph{J. Anim. Ecol.}, 59:\penalty0 1147--1164, 1990.

\bibitem[Levin(1992)]{levin:1992}
S.~A. Levin.
\newblock The problem of pattern and scale in ecology: The {Robert H. MacArthur
  Award} lecture.
\newblock \emph{Ecology}, 73:\penalty0 1943--1967, 1992.

\bibitem[Stadler et~al.(2001)Stadler, Stadler, Wagner, and
  Fontana]{stadler:2001}
B.~M.~R. Stadler, P.~F. Stadler, G.~P. Wagner, and W.~Fontana.
\newblock The topology of the possible: Formal spaces underlying patterns of
  evolutionary change.
\newblock \emph{J.\ Theor.\ Biol.}, 213:\penalty0 241--274, 2001.

\bibitem[Huynen et~al.(1996)Huynen, Stadler, and Fontana]{huynen:1996}
M.~A. Huynen, P.~F. Stadler, and W.~Fontana.
\newblock Smoothness within ruggedness: The role of neutrality in adaptation.
\newblock \emph{Proc.\ Natl.\ Acad.\ Sci.\ USA}, 93:\penalty0 397--401, 1996.

\bibitem[Aguirre and Manrubia(2015)]{aguirre:2015}
J.~Aguirre and S.~Manrubia.
\newblock Tipping points and early warning signals in the genomic composition
  of populations induced by environmental changes.
\newblock \emph{Sci.\ Rep.}, 5:\penalty0 9664, 2015.

\bibitem[Koelle et~al.(2006)Koelle, Cobey, Grenfell, and Pascual]{koelle:2006}
K.~Koelle, S.~Cobey, B.~Grenfell, and M.~Pascual.
\newblock {Epochal evolution shapes the phylodynamics of interpandemic
  influenza {A (H3N2)} in humans}.
\newblock \emph{Science}, 314:\penalty0 1898--1903, 2006.

\bibitem[Wolf et~al.(2006)Wolf, Viboud, Holmes, Koonin, and Lipman]{wolf:2006}
Y.~I. Wolf, C.~Viboud, E.~C. Holmes, E.~V. Koonin, and D.~J. Lipman.
\newblock Long intervals of stasis punctuated by bursts of positive selection
  in the seasonal evolution of influenza {A} virus.
\newblock \emph{Biol.\ Direct}, 1:\penalty0 34, 2006.

\bibitem[Wright(1931)]{wright:1931}
S.~Wright.
\newblock Evolution in {Mendelian} populations.
\newblock \emph{Genetics}, 16:\penalty0 97--159, 1931.

\bibitem[Svensson and Calsbeek(2012)]{svensson:2012}
E.~I. Svensson and R.~Calsbeek.
\newblock \emph{{The adaptive landscape in evolutionary biology}}.
\newblock Oxford University Press, 2012.

\bibitem[Wagner(2011)]{wagner:2011}
A.~Wagner.
\newblock \emph{The origins of evolutionary innovations}.
\newblock Oxford University Press, 2011.

\bibitem[Conant and Wolfe(2008)]{conant:2008}
G.~C. Conant and K.~H. Wolfe.
\newblock Turning a hobby into a job: How duplicated genes find new functions.
\newblock \emph{Nat.\ Rev.\ Genet.}, 9:\penalty0 938--950, 2008.

\bibitem[Catal\'an et~al.(2017)Catal\'an, Arias, Cuesta, and
  Manrubia]{catalan:2017}
P.~Catal\'an, C.~F. Arias, J.~A. Cuesta, and S.~Manrubia.
\newblock Adaptive multiscapes: {An} up-to-date metaphor to visualize molecular
  adaptation.
\newblock \emph{Biol. Direct}, 12:\penalty0 7, 2017.

\bibitem[Hinkley et~al.(2011)Hinkley, Martins, Chappey, Haddad, Stawiski,
  Whitcomb, Petropoulos, and Bonhoeffer]{hinkley:2011}
T.~Hinkley, J.~Martins, C.~Chappey, M.~Haddad, E.~Stawiski, J.~M. Whitcomb,
  C.~J. Petropoulos, and S.~Bonhoeffer.
\newblock A systems analysis of mutational effects in {HIV}-1 protease and
  reverse transcriptase.
\newblock \emph{Nat.\ Genet.}, 43:\penalty0 487--489, 2011.

\bibitem[Payne and Wagner(2014{\natexlab{a}})]{payne:2014b}
J.~L. Payne and A.~Wagner.
\newblock The robustness and evolvability of transcription factor binding
  sites.
\newblock \emph{Science}, 343:\penalty0 875--877, 2014{\natexlab{a}}.

\bibitem[Steinberg and Ostermeier(2016)]{steinberg:2016}
B.~Steinberg and M.~Ostermeier.
\newblock Environmental changes bridge evolutionary valleys.
\newblock \emph{Sci.\ Adv.}, 2:\penalty0 e1500921, 2016.

\bibitem[Waddington(1953)]{waddington:1953}
C.~H. Waddington.
\newblock Genetic assimilation of an acquired character.
\newblock \emph{Evolution}, 7:\penalty0 118--126, 1953.

\bibitem[Kimura(1968)]{kimura:1968}
M.~Kimura.
\newblock Evolutionary rate at the molecular level.
\newblock \emph{Nature}, 217:\penalty0 624--626, 1968.

\bibitem[Kimura(1984)]{kimura:1984}
M.~Kimura.
\newblock \emph{The Neutral Theory of Molecular Evolution}.
\newblock Cambridge University Press, 1984.

\bibitem[{Maynard Smith}(1970)]{maynard-smith:1970}
J.~{Maynard Smith}.
\newblock Natural selection and the concept of a protein space.
\newblock \emph{Nature}, 225:\penalty0 563--564, 1970.

\bibitem[Schuster et~al.(1994)Schuster, Fontana, Stadler, and
  Hofacker]{schuster:1994}
P.~Schuster, W.~Fontana, P.~F. Stadler, and I.~L. Hofacker.
\newblock From sequences to shapes and back: A case study in {RNA} secondary
  structures.
\newblock \emph{Proc. Roy. Soc. London B}, 255:\penalty0 279--284, 1994.

\bibitem[Bastolla et~al.(2003)Bastolla, Porto, Roman, and
  Vendruscolo]{bastolla:2003}
U.~Bastolla, M.~Porto, H.~E. Roman, and M.~Vendruscolo.
\newblock Connectivity of neutral networks, overdispersion, and structural
  conservation in protein evolution.
\newblock \emph{J. Mol. Biol.}, 56:\penalty0 243--254, 2003.

\bibitem[Ciliberti et~al.(2007)Ciliberti, Martin, and Wagner]{ciliberti:2007}
S.~Ciliberti, O.~C. Martin, and A.~Wagner.
\newblock Innovation and robustness in complex regulatory gene networks.
\newblock \emph{Proc.\ Natl.\ Acad.\ Sci.\ USA}, 104:\penalty0 13595--13596,
  2007.

\bibitem[{Matias Rodrigues} and Wagner(2011)]{matias-rodrigues:2011}
J.~F. {Matias Rodrigues} and A.~Wagner.
\newblock Genotype networks, innovation, and robustness in sulfur metabolism.
\newblock \emph{BMC Syst.\ Biol.}, 5:\penalty0 39, 2011.

\bibitem[Schultes and Bartel(2000)]{schultes:2000}
E.~A. Schultes and D.~P. Bartel.
\newblock One sequence, two ribozymes: implications for the emergence of new
  ribozyme folds.
\newblock \emph{Science}, 289:\penalty0 448--452, 2000.

\bibitem[Bloom et~al.(2007{\natexlab{a}})Bloom, Romero, Lu, and
  Arnold]{bloom:2007b}
J.~D. Bloom, P.~A. Romero, Z.~Lu, and F.~H. Arnold.
\newblock Neutral genetic drift can alter promiscuous protein functions,
  potentially aiding functional evolution.
\newblock \emph{Biol.\ Dir.}, 2:\penalty0 17, 2007{\natexlab{a}}.

\bibitem[Aguilar-Rodr{\'\i}guez et~al.(2017)Aguilar-Rodr{\'\i}guez, Payne, and
  Wagner]{aguilar:2017}
J.~Aguilar-Rodr{\'\i}guez, J.~L. Payne, and A.~Wagner.
\newblock A thousand empirical adaptive landscapes and their navigability.
\newblock \emph{Nature Ecol. Evol.}, 1:\penalty0 45, 2017.

\bibitem[Ohta(1973)]{ohta:1973}
T.~Ohta.
\newblock Slightly deleterious mutant substitutions in evolution.
\newblock \emph{Nature}, 246:\penalty0 96--98, 1973.

\bibitem[Eigen(1971)]{eigen:1971}
M.~Eigen.
\newblock Selforganization of matter and the evolution of biological
  macromolecules.
\newblock \emph{Naturwissenschaften}, 58:\penalty0 465--523, 1971.

\bibitem[Domingo(2006)]{domingo:2006}
E.~Domingo, editor.
\newblock \emph{Quasispecies: Concept and Implications for Virology}.
\newblock Springer, Berlin, 2006.

\bibitem[Woo and Reifman(2012)]{woo:2012}
H.-J. Woo and J.~Reifman.
\newblock A quantitative quasispecies theory-based model of virus escape
  mutation under immune selection.
\newblock \emph{Proc.\ Natl.\ Acad.\ Sci.\ USA}, 109:\penalty0 12980--12985,
  2012.

\bibitem[Hofacker et~al.(1994)Hofacker, Fontana, Stadler, Bonhoeffer, Tacker,
  and Schuster]{hofacker:1994}
I.~L. Hofacker, W.~Fontana, P.~F. Stadler, L.~S. Bonhoeffer, M.~Tacker, and
  P.~Schuster.
\newblock Fast folding and comparison of {RNA} secondary structures.
\newblock \emph{Monatshefte f. {C}hemie}, 125:\penalty0 167--188, 1994.

\bibitem[Gr{\"u}ner et~al.(1996{\natexlab{a}})Gr{\"u}ner, Giegerich,
  Strothmann, Reidys, Weber, Hofacker, Stadler, and Schuster]{gruner:1996}
W.~Gr{\"u}ner, R.~Giegerich, D.~Strothmann, C.~Reidys, J.~Weber, I.~L.
  Hofacker, P.~F. Stadler, and P.~Schuster.
\newblock Analysis of {RNA} sequence structure maps by exhaustive enumeration.
  {I. Neutral} networks.
\newblock \emph{Monatshefte f. Chemie}, 127:\penalty0 355--374,
  1996{\natexlab{a}}.

\bibitem[Fontana and Schuster(1998{\natexlab{a}})]{fontana:1998}
W.~Fontana and P.~Schuster.
\newblock Continuity in evolution: on the nature of transitions.
\newblock \emph{Science}, 280:\penalty0 1451--1455, 1998{\natexlab{a}}.

\bibitem[Cowperthwaite et~al.(2008)Cowperthwaite, Economo, Harcombe, Miller,
  and Meyers]{cowperthwaite:2008}
M.~C. Cowperthwaite, E.~P. Economo, W.~R. Harcombe, E.~L. Miller, and L.~A.
  Meyers.
\newblock The ascent of the abundant: How mutational networks constrain
  evolution.
\newblock \emph{PLoS Comput.\ Biol.}, 4:\penalty0 e1000110, 2008.

\bibitem[J\"org et~al.(2008)J\"org, Martin, and Wagner]{jorg:2008}
T.~J\"org, O.~C. Martin, and A.~Wagner.
\newblock Neutral network sizes of biological {RNA} molecules can be computed
  and are not atypically small.
\newblock \emph{BMC Bioinformatics}, 9:\penalty0 464, 2008.

\bibitem[Aguirre et~al.(2011)Aguirre, Buld\'u, Stich, and
  Manrubia]{aguirre:2011}
J.~Aguirre, J.~M. Buld\'u, M.~Stich, and S.~C. Manrubia.
\newblock Topological structure of the space of phenotypes: the case of {RNA}
  neutral networks.
\newblock \emph{PLoS ONE}, 6:\penalty0 e26324, 2011.

\bibitem[Dingle et~al.(2015)Dingle, Schaper, and Louis]{dingle:2015}
K.~Dingle, S.~Schaper, and A.~A. Louis.
\newblock The structure of the genotype{-}phenotype map strongly constrains the
  evolution of non-coding {RNA}.
\newblock \emph{J.\ Roy.\ Soc.\ Interface}, 5:\penalty0 20150053, 2015.

\bibitem[Lau and Dill(1989)]{lau:1989}
K.~F. Lau and K.~A. Dill.
\newblock A lattice statistical mechanics model of the conformational and
  sequence spaces of proteins.
\newblock \emph{Macromolecules}, 22:\penalty0 3986--3997, 1989.

\bibitem[Lipman and Wilbur(1991)]{lipman:1991}
D.~J. Lipman and W.~J. Wilbur.
\newblock Modelling neutral and selective evolution of protein folding.
\newblock \emph{Proc. Roy. Soc. London B}, 245:\penalty0 7--11, 1991.

\bibitem[Li et~al.(1996)Li, Helling, Tang, and Wingreen]{li:1996}
H.~Li, R.~Helling, C.~Tang, and N.~Wingreen.
\newblock Emergence of preferred structures in a simple model of protein
  folding.
\newblock \emph{Science}, 273:\penalty0 666--669, 1996.

\bibitem[Bornberg-Bauer(1997)]{bornberg-bauer:1997}
E.~Bornberg-Bauer.
\newblock How are model protein structures distributed in sequence space?
\newblock \emph{Biophys. J.}, 73:\penalty0 2393, 1997.

\bibitem[Irb\"ack and Troein(2002)]{irback:2002}
A.~Irb\"ack and C.~Troein.
\newblock Enumerating designing sequences in the {HP} model.
\newblock \emph{J.\ Biol.\ Phys.}, 28:\penalty0 1--15, 2002.

\bibitem[Crombach(2017)]{crombach:2017}
A.~Crombach.
\newblock \emph{Modelling the Evolution of Dynamic Regulatory Networks: Some
  Critical Insights}, pages 257--273.
\newblock Springer International Publishing, Cham, 2017.

\bibitem[Payne et~al.(2014)Payne, Moore, and Wagner]{payne:2014a}
J.~L. Payne, J.~H. Moore, and A.~Wagner.
\newblock Robustness, evolvability, and the logic of genetic regulation.
\newblock \emph{Artif. {L}ife}, 20:\penalty0 111--126, 2014.

\bibitem[{Matias Rodrigues} and Wagner(2009)]{matias-rodrigues:2009}
J.~F. {Matias Rodrigues} and A.~Wagner.
\newblock Evolutionary plasticity and innovations in complex metabolic reaction
  networks.
\newblock \emph{PLoS Comp. Biol.}, 5:\penalty0 e1000613, 2009.

\bibitem[Barve and Wagner(2013)]{barve:2013}
A.~Barve and A.~Wagner.
\newblock A latent capacity for evolutionary innovation through exaptation in
  metabolic systems.
\newblock \emph{Nature}, 500:\penalty0 203--206, 2013.

\bibitem[Hosseini et~al.(2015)Hosseini, Barve, and Wagner]{hosseini:2015}
S.-R. Hosseini, A.~Barve, and A.~Wagner.
\newblock Exhaustive analysis of a genotype space comprising $10^{15}$ central
  carbon metabolisms reveals an organization conducive to metabolic innovation.
\newblock \emph{PLoS {C}omput. {B}iol.}, 11:\penalty0 e1004329, 2015.

\bibitem[Ib\'a{\~n}ez-Marcelo and Alarc\'on(2104)]{ibanez:2014}
E.~Ib\'a{\~n}ez-Marcelo and T.~Alarc\'on.
\newblock The topology of robustness and evolvability in evolutionary systems
  with genotype-phenotype map.
\newblock \emph{J.\ Theor.\ Biol.}, 356:\penalty0 144--162, 2104.

\bibitem[Johnston et~al.(2011)Johnston, Ahnert, Doye, and Louis]{johnston:2011}
I.~G. Johnston, S.~E. Ahnert, J.~P. Doye, and A.~A. Louis.
\newblock Evolutionary dynamics in a simple model of self-assembly.
\newblock \emph{Phys.\ Rev.\ E}, 83:\penalty0 066105, 2011.

\bibitem[Greenbury et~al.(2014)Greenbury, Johnston, Louis, and
  Ahnert]{greenbury:2014}
S.~F. Greenbury, I.~G. Johnston, A.~A. Louis, and S.~E. Ahnert.
\newblock A tractable {genotype-phenotype map} for the self-assembly of protein
  quaternary structure.
\newblock \emph{J.\ R.\ Soc.\ Interface}, 11:\penalty0 20140249, 2014.

\bibitem[Arias et~al.(2014)Arias, Catal{\'a}n, Manrubia, and
  Cuesta]{arias:2014}
C.~F. Arias, P.~Catal{\'a}n, S.~Manrubia, and J.~A. Cuesta.
\newblock toy{LIFE}: a computational framework to study the multi-level
  organisation of the genotype-phenotype map.
\newblock \emph{Sci. {R}ep.}, 4:\penalty0 7549, 2014.

\bibitem[Catal\'an et~al.(2018)Catal\'an, Wagner, Manrubia, and
  Cuesta]{catalan:2018}
P.~Catal\'an, A.~Wagner, S.~Manrubia, and J.~A. Cuesta.
\newblock Adding levels of complexity enhances robustness and evolvability in a
  multi-level genotype-phenotype map.
\newblock \emph{J.\ Roy.\ Soc.\ Interface}, 15:\penalty0 20170516, 2018.

\bibitem[Greenbury and Ahnert(2015)]{greenbury:2015}
S.~Greenbury and S.~Ahnert.
\newblock The organization of biological sequences into constrained and
  unconstrained parts determines fundamental properties of genotype--phenotype
  maps.
\newblock \emph{J.\ R.\ Soc.\ Interface}, 12:\penalty0 20150724, 2015.

\bibitem[{Manrubia} and {Cuesta}(2017)]{manrubia:2017}
S.~{Manrubia} and J.~A. {Cuesta}.
\newblock {Distribution of genotype network sizes in sequence-to-structure
  genotype-phenotype maps}.
\newblock \emph{J.\ R.\ Soc.\ Interface}, 14:\penalty0 20160976, 2017.

\bibitem[Ahnert(2017)]{ahnert:2017}
S.~E. Ahnert.
\newblock Structural properties of genotype-phenotype maps.
\newblock \emph{J.\ R.\ Soc.\ Interface}, 14:\penalty0 20170275, 2017.

\bibitem[Ferrada and Wagner(2012)]{ferrada:2012}
E.~Ferrada and A.~Wagner.
\newblock A comparison of genotype-phenotype maps for {RNA} and proteins.
\newblock \emph{Biophys.\ J.}, 102:\penalty0 1916--1925, 2012.

\bibitem[Khatri et~al.(2009)Khatri, McLeish, and Sear]{khatri:2009}
B.~S. Khatri, T.~C.~B. McLeish, and R.~P. Sear.
\newblock Statistical mechanics of convergent evolution in spatial patterning.
\newblock \emph{Proceedings of the National Academy of Sciences}, 106\penalty0
  (24):\penalty0 9564--9569, 2009.

\bibitem[Fontana and Schuster(1998{\natexlab{b}})]{fontana:1998b}
W.~Fontana and P.~Schuster.
\newblock Shaping space: The possible and the attainable in {RNA}
  genotype-phenotype mapping.
\newblock \emph{J.\ Theor.\ Biol.}, 194:\penalty0 491--515, 1998{\natexlab{b}}.

\bibitem[Schaper and Louis(2014)]{schaper:2014}
S.~Schaper and A.~A. Louis.
\newblock The arrival of the frequent: How bias in genotype-phenotype maps can
  steer populations to local optima.
\newblock \emph{PLoS ONE}, 9:\penalty0 e86635, 2014.

\bibitem[{Manrubia} and {Cuesta}(2015)]{manrubia:2015}
S.~{Manrubia} and J.~A. {Cuesta}.
\newblock {Evolution on genotype networks accelerates the ticking rate of the
  molecular clock}.
\newblock \emph{J.\ R.\ Soc.\ Interface}, 12:\penalty0 20141010, 2015.

\bibitem[Greenbury et~al.(2016)Greenbury, Schaper, Ahnert, and
  Louis]{greenbury:2016}
S.~F. Greenbury, S.~Schaper, S.~E. Ahnert, and A.~A. Louis.
\newblock Genetic correlations greatly increase mutational robustness and can
  both reduce and enhance evolvability.
\newblock \emph{PLoS Comput.\ Biol.}, 12:\penalty0 e1004773, 2016.

\bibitem[Bornberg-Bauer and Chan(1999)]{bornberg-bauer:1999}
E.~Bornberg-Bauer and H.~S. Chan.
\newblock Modeling evolutionary landscapes: Mutational stability, topology, and
  superfunnels in sequence space.
\newblock \emph{Proc. Natl. Acad. Sci. USA}, 96:\penalty0 10689--10694, 1999.

\bibitem[Wuchty et~al.(1999)Wuchty, Fontana, Hofacker, and
  Schuster]{wuchty:1999}
S.~Wuchty, W.~Fontana, I.~L. Hofacker, and P.~Schuster.
\newblock Complete suboptimal folding of {RNA} and the stability of secondary
  structures.
\newblock \emph{Biopolymers}, 49:\penalty0 145--165, 1999.

\bibitem[Ancel and Fontana(2000)]{ancel:2000}
L.~W. Ancel and W.~Fontana.
\newblock Plasticity, evolvability, and modularity in {RNA}.
\newblock \emph{J.\ Exp.\ Zool.}, 288:\penalty0 242--283, 2000.

\bibitem[Gr{\"u}ner et~al.(1996{\natexlab{b}})Gr{\"u}ner, Giegerich,
  Strothmann, Reidys, Weber, Hofacker, Stadler, and Schuster]{gruner:1996b}
W.~Gr{\"u}ner, R.~Giegerich, D.~Strothmann, C.~Reidys, J.~Weber, I.~L.
  Hofacker, P.~F. Stadler, and P.~Schuster.
\newblock Analysis of {RNA} sequence structure maps by exhaustive enumeration
  {II. Structures} of neutral networks and shape space covering.
\newblock \emph{Monatshefte f. {C}hemie}, 127:\penalty0 375--389,
  1996{\natexlab{b}}.

\bibitem[Kivel{\"a} et~al.(2010)Kivel{\"a}, Arenas, Barthelemy, Gleeson,
  Moreno, and Porter]{kivela:2014}
M.~Kivel{\"a}, A.~Arenas, M.~Barthelemy, J.~P. Gleeson, Y.~Moreno, and M.~A.
  Porter.
\newblock Multilayer networks.
\newblock \emph{J. Comp. Net.}, 2:\penalty0 203--271, 2010.

\bibitem[Boccaletti et~al.(2014)Boccaletti, Bianconi, Criado, del Genio,
  G{\'o}mez-Garde{\~n}es, Romance, Sendi{\~n}a-Nadal, Wang, and
  Zanin]{boccaletti:2014}
S.~Boccaletti, G.~Bianconi, R.~Criado, C.~del Genio, J.~G{\'o}mez-Garde{\~n}es,
  M.~Romance, I.~Sendi{\~n}a-Nadal, Z.~Wang, and M.~Zanin.
\newblock {The structure and dynamics of multilayer networks}.
\newblock \emph{Phys.\ Rep.}, 544:\penalty0 1--122, 2014.

\bibitem[Manrubia(2012)]{manrubia:2012}
S.~C. Manrubia.
\newblock {Modelling viral evolution and adaptation: challenges and rewards}.
\newblock \emph{Curr.\ Opin.\ Virol.}, 2:\penalty0 531--537, 2012.

\bibitem[Li et~al.(2002)Li, Tang, and Wingreen]{li:2002}
H.~Li, C.~Tang, and N.~S. Wingreen.
\newblock Designability of protein structures: {A} lattice-model study using
  the {Miyazawa-Jernigan} matrix.
\newblock \emph{Proteins}, 49:\penalty0 403--412, 2002.

\bibitem[Stich et~al.(2010)Stich, L\'azaro, and Manrubia]{stich:2010}
M.~Stich, E.~L\'azaro, and S.~C. Manrubia.
\newblock Phenotypic effect of mutations in evolving populations of {RNA}
  molecules.
\newblock \emph{BMC Evol.\ Biol.}, 10:\penalty0 46, 2010.

\bibitem[Poelwijk et~al.(2007)Poelwijk, Kiviet, Weinreich, and
  Tans]{poelwijk:2007}
F.~J. Poelwijk, D.~J. Kiviet, D.~M. Weinreich, and S.~J. Tans.
\newblock Empirical fitness landscapes reveal accessible evolutionary paths.
\newblock \emph{Nature}, 445:\penalty0 383, 2007.

\bibitem[Jim{\'e}nez et~al.(2013)Jim{\'e}nez, Xulvi-Brunet, Campbell,
  Turk-MacLeod, and Chen]{jimenez:2013}
J.~I. Jim{\'e}nez, R.~Xulvi-Brunet, G.~W. Campbell, R.~Turk-MacLeod, and I.~A.
  Chen.
\newblock Comprehensive experimental fitness landscape and evolutionary network
  for small {RNA}.
\newblock \emph{Proc.\ Natl.\ Acad.\ Sci.\ USA}, 110:\penalty0 14984--14989,
  2013.

\bibitem[De~Visser and Krug(2014)]{devisser:2014}
J.~A.~G. De~Visser and J.~Krug.
\newblock Empirical fitness landscapes and the predictability of evolution.
\newblock \emph{Nat. Rev. Genet.}, 15:\penalty0 480, 2014.

\bibitem[De~Vos et~al.(2015)De~Vos, Dawid, Sunderlikova, and Tans]{devos:2015}
M.~G. De~Vos, A.~Dawid, V.~Sunderlikova, and S.~J. Tans.
\newblock Breaking evolutionary constraint with a tradeoff ratchet.
\newblock \emph{Proc.\ Natl.\ Acad.\ Sci.\ USA}, 112:\penalty0 14906--14911,
  2015.

\bibitem[Firnberg et~al.(2014)Firnberg, Labonte, Gray, and
  Ostermeier]{firnberg:2014}
E.~Firnberg, J.~W. Labonte, J.~J. Gray, and M.~Ostermeier.
\newblock A comprehensive, high-resolution map of a gene’s fitness landscape.
\newblock \emph{Mol.\ Biol.\ Evol.}, 31:\penalty0 1581--1592, 2014.

\bibitem[Lauring and Andino(2011)]{lauring:2011}
A.~S. Lauring and R.~Andino.
\newblock Exploring the fitness landscape of an {RNA} virus by using a
  universal barcode microarray.
\newblock \emph{J.\ Virol.}, 85:\penalty0 3780--3791, 2011.

\bibitem[Kouyos et~al.(2012)Kouyos, Leventhal, Hinkley, Haddad, Whitcomb,
  Petropoulos, and Bonhoeffer]{kouyos:2012}
R.~D. Kouyos, G.~E. Leventhal, T.~Hinkley, M.~Haddad, J.~M. Whitcomb, C.~J.
  Petropoulos, and S.~Bonhoeffer.
\newblock Exploring the complexity of the {HIV}-1 fitness landscape.
\newblock \emph{PLoS Genetics}, 8:\penalty0 e1002551, 2012.

\bibitem[Bank et~al.(2016)Bank, Matuszewski, Hietpas, and Jensen]{bank:2016}
C.~Bank, S.~Matuszewski, R.~T. Hietpas, and J.~D. Jensen.
\newblock On the (un)predictability of a large intragenic fitness landscape.
\newblock \emph{Proc.\ Natl.\ Acad.\ Sci.\ USA}, 113:\penalty0 14085--14090,
  2016.

\bibitem[Zagorski et~al.(2016)Zagorski, Burda, and Waclaw]{zagorski:2016}
M.~Zagorski, Z.~Burda, and B.~Waclaw.
\newblock Beyond the hypercube: Evolutionary accessibility of fitness
  landscapes with realistic mutational networks.
\newblock \emph{PLoS Comp.\ Biol.}, 12:\penalty0 e1005218, 2016.

\bibitem[Kauffman and Levin(1987)]{kauffman:1987}
S.~Kauffman and S.~Levin.
\newblock Towards a general theory of adaptive walks on rugged landscapes.
\newblock \emph{J.\ Theor.\ Biol.}, 128:\penalty0 11 -- 45, 1987.

\bibitem[{\O}stman and Adami(2014)]{ostman:2014}
B.~{\O}stman and C.~Adami.
\newblock Predicting evolution and visualizing high-dimensional fitness
  landscapes.
\newblock In H.~Richter and A.~Engelbrecht, editors, \emph{Recent Advances in
  the Theory and Application of Fitness Landscapes}, pages 509--526. Springer
  Berlin Heidelberg, Berlin, Heidelberg, 2014.

\bibitem[Yubero et~al.(2017)Yubero, Manrubia, and Aguirre.]{yubero:2017}
P.~Yubero, S.~Manrubia, and J.~Aguirre.
\newblock The space of genotypes is a network of networks: implications for
  evolutionary and extinction dynamics.
\newblock \emph{Sci.\ Rep.}, 7:\penalty0 13813, 2017.

\bibitem[Wilke(2001)]{wilke:2001}
C.~O. Wilke.
\newblock Adaptive evolution on neutral networks.
\newblock \emph{Bull.\ Math.\ Biol.}, 63:\penalty0 715--730, 2001.

\bibitem[Aguirre et~al.(2009)Aguirre, Buld\'u, and Manrubia]{aguirre:2009}
J.~Aguirre, J.~M. Buld\'u, and S.~C. Manrubia.
\newblock Evolutionary dynamics on networks of selectively neutral genotypes:
  Effects of topology and sequence stability.
\newblock \emph{Phys.\ Rev.\ E}, 80:\penalty0 066112, 2009.

\bibitem[com()]{comment1}
In~\cite{aguirre:2015,yubero:2017} a slightly different form of this equation
  was used, with the matrix product ({\bf F G}) instead of ({\bf G F}) in the
  second term on the right. While the former can be interpreted as a filtering
  criteria for the stability of mutants (fitness is applied once mutation has
  occurred), the latter represents more precisely the evolutionary process here
  described (fitness affects replication rates and mutation occurs
  concomitantly). The qualitative results of both expressions are equivalent,
  in practice only differing in the specific moment when the population state
  is measured.

\bibitem[Bocaletti et~al.(2006)Bocaletti, Latora, Moreno, Chavez, and
  Hwang]{boccaletti:2006}
S.~Bocaletti, V.~Latora, Y.~Moreno, M.~Chavez, and D.~U. Hwang.
\newblock Complex networks: structure and dynamics.
\newblock \emph{Phys.\ Rep.}, 424:\penalty0 175--308, 2006.

\bibitem[Reeves et~al.(2016)Reeves, Farr, Blundell, Gallagher, and
  Fink]{reeves:2016}
T.~Reeves, R.~S. Farr, J.~Blundell, A.~Gallagher, and T.~M.~A. Fink.
\newblock Eigenvalues of neutral networks: Interpolating between hypercubes.
\newblock \emph{Discrete Mathematics}, 339:\penalty0 1283--1290, 2016.

\bibitem[{van Nimwegen} et~al.(1999){van Nimwegen}, Crutchfield, and
  Huynen]{nimwegen:1999}
E.~{van Nimwegen}, J.~P. Crutchfield, and M.~Huynen.
\newblock Neutral evolution of mutational robustness.
\newblock \emph{Proc.\ Natl.\ Acad.\ Sci.\ USA}, 96:\penalty0 9716--9720, 1999.

\bibitem[Wagner(2008)]{wagner:2008}
A.~Wagner.
\newblock Robustness and evolvability: a paradox resolved.
\newblock \emph{Proc.\ R.\ Soc.\ Lond.\ B}, 275:\penalty0 91--100, 2008.

\bibitem[Draghi et~al.(2010)Draghi, Parsons, Wagner, and Plotkin]{draghi:2010}
J.~A. Draghi, T.~L. Parsons, G.~P. Wagner, and J.~B. Plotkin.
\newblock {Mutational robustness can facilitate adaptation}.
\newblock \emph{Nature}, 463:\penalty0 353--355, 2010.

\bibitem[Reidys et~al.(1997)Reidys, Stadler, and Schuster]{reidys:1997}
C.~Reidys, P.~F. Stadler, and P.~Schuster.
\newblock Generic properties of combinatory maps: Neutral networks of {RNA}
  secondary structures.
\newblock \emph{Bull.\ Math.\ Biol.}, 59:\penalty0 339--397, Mar 1997.

\bibitem[Fontana(2002)]{fontana:2002}
W.~Fontana.
\newblock Modelling `evo-devo' with {RNA}.
\newblock \emph{BioEssays}, 24:\penalty0 1164--1177, 2002.

\bibitem[Zuckerkandl and Pauling(1965)]{zuckerkandl:1965}
E.~Zuckerkandl and L.~Pauling.
\newblock Evolutionary divergence and convergence in proteins.
\newblock In V.~Bryson and H.~Vogel, editors, \emph{Evolving genes and
  proteins}, pages 97--166. Academic Press, New York, NY, 1965.

\bibitem[Takahata(1987)]{takahata:1987}
N.~Takahata.
\newblock On the overdispersed molecular clock.
\newblock \emph{Genetics}, 116:\penalty0 169--179, 1987.

\bibitem[Aguirre et~al.(2013)Aguirre, Papo, and Buld{\'u}]{aguirre:2013}
J.~Aguirre, D.~Papo, and J.~M. Buld{\'u}.
\newblock Successful strategies for competing networks.
\newblock \emph{Nature Phys.}, 9:\penalty0 230--234, 2013.

\bibitem[Buld{\'u} et~al.(2016)Buld{\'u}, Sevilla-Escoboza, Aguirre, Papo, and
  Guti{\'e}rrez]{buldu:2016}
J.~M. Buld{\'u}, R.~Sevilla-Escoboza, J.~Aguirre, D.~Papo, and
  R.~Guti{\'e}rrez.
\newblock Interconnecting networks: The role of connector links.
\newblock In A.~Garas, editor, \emph{Interconnected Networks}, pages 61--77.
  Springer International Publishing, 2016.

\bibitem[Iranzo et~al.(2016{\natexlab{a}})Iranzo, Buld{\'u}, and
  Aguirre]{iranzo:2016}
J.~Iranzo, J.~M. Buld{\'u}, and J.~Aguirre.
\newblock {Competition among networks highlights the power of the weak}.
\newblock \emph{Nature Comm.}, 7:\penalty0 13273, 2016{\natexlab{a}}.

\bibitem[Newman(2010)]{newman:2010}
M.~E.~J. Newman.
\newblock \emph{Networks: An introduction}.
\newblock Oxford University Press, New York, 2010.

\bibitem[Langville.(2006)]{langville:2006}
C.~Langville., A.N. \&~Meyer.
\newblock \emph{Google's PageRank and Beyond: The Science of Search Engine
  Rankings}.
\newblock Princeton University Press, 2006.

\bibitem[Senanayake et~al.(2015)Senanayake, Piraveenan, and
  Zomaya]{senanayake:2015}
U.~Senanayake, M.~Piraveenan, and A.~Zomaya.
\newblock The pagerank-index: Going beyond citation counts in quantifying
  scientific impact of researchers.
\newblock \emph{PLoS ONE}, 10:\penalty0 1--34, 2015.

\bibitem[Bergstrom(2007)]{bergstrom:2007}
C.~Bergstrom.
\newblock Eigenfactor: Measuring the value and prestige of scholarly journals.
\newblock \emph{C{\&}RL News}, 68:\penalty0 314--316, 2007.

\bibitem[Seary and Richards(2003)]{seary:2003}
A.~J. Seary and W.~D. Richards.
\newblock Spectral methods for analyzing and visualizing networks: an
  introduction.
\newblock In \emph{Dynamic Social Network Modeling and Analysis. Workshop
  Summary and Papers}, pages 209--228. The National Academies Press,
  Washington, D. C., 2003.

\bibitem[Lohmann et~al.(2010)Lohmann, Margulies, Horstmann, Pleger, Lepsien,
  Goldhahn, Schloegl, Stumvoll, Villringer, and Turner]{lohmann:2010}
G.~Lohmann, D.~S. Margulies, A.~Horstmann, B.~Pleger, J.~Lepsien, D.~Goldhahn,
  H.~Schloegl, M.~Stumvoll, A.~Villringer, and R.~Turner.
\newblock Eigenvector centrality mapping for analyzing connectivity patterns in
  {fMRI} data of the human brain.
\newblock \emph{PLoS ONE}, 5:\penalty0 1--8, 2010.

\bibitem[{van Nimwegen} and Crutchfield(2000{\natexlab{a}})]{nimwegen:2000}
E.~{van Nimwegen} and J.~P. Crutchfield.
\newblock Metastable evolutionary dynamics: crossing fitness barriers or
  escaping via neutral paths?
\newblock \emph{Bull.\ Math.\ Biol.}, 62:\penalty0 799--848,
  2000{\natexlab{a}}.

\bibitem[Wilke et~al.(2001)Wilke, Wang, Ofria, Lenski, and Adami]{wilke:2001b}
C.~O. Wilke, J.~L. Wang, C.~Ofria, R.~E. Lenski, and C.~Adami.
\newblock Evolution of digital organisms at high mutation rates leads to
  survival of the flattest.
\newblock \emph{Nature}, 412:\penalty0 331--333, 2001.

\bibitem[Codo{\~n}er et~al.(2006)Codo{\~n}er, Dar{\'o}s, Sol{\'e}, and
  Elena]{codoner:2006}
F.~M. Codo{\~n}er, J.-A. Dar{\'o}s, R.~V. Sol{\'e}, and S.~F. Elena.
\newblock The fittest versus the flattest: experimental confirmation of the
  quasispecies effect with subviral pathogens.
\newblock \emph{PLoS Pathog.}, 2:\penalty0 e136, 2006.

\bibitem[Wilke and Adami(2003)]{wilke:2003}
C.~O. Wilke and C.~Adami.
\newblock Evolution of mutational robustness.
\newblock \emph{Mut.\ Res.}, 522:\penalty0 3--11, 2003.

\bibitem[{van Nimwegen} and Crutchfield(2000{\natexlab{b}})]{nimwegen:2000b}
E.~{van Nimwegen} and J.~P. Crutchfield.
\newblock Optimizing epochal evolutionary search: Population-size independent
  theory.
\newblock \emph{Comput.\ Methods Appl.\ Mech.\ Eng.}, 186:\penalty0 171--194,
  2000{\natexlab{b}}.

\bibitem[{van Nimwegen} and Crutchfield(2001)]{nimwegen:2001}
E.~{van Nimwegen} and J.~P. Crutchfield.
\newblock Optimizing epochal evolutionary search: Population-size dependent
  theory.
\newblock \emph{Mach. Learn.}, 45:\penalty0 77--114, 2001.

\bibitem[Fuentes-Hernandez et~al.(2015)Fuentes-Hernandez, Plucain, Gori,
  Pena-Miller, Reding, Jansen, Schulenburg, Gudelj, and
  Beardmore]{fuentes:2015}
A.~Fuentes-Hernandez, J.~Plucain, F.~Gori, R.~Pena-Miller, C.~Reding,
  G.~Jansen, H.~Schulenburg, I.~Gudelj, and R.~Beardmore.
\newblock Using a sequential regimen to eliminate bacteria at sublethal
  antibiotic dosages.
\newblock \emph{PLoS Biol.}, 13:\penalty0 e1002104, 2015.

\bibitem[Gavrilets(2004)]{gavrilets:2004}
S.~Gavrilets.
\newblock \emph{Fitness Landscapes and the Origin of Species}.
\newblock Princeton University Press, Princeton, 2004.

\bibitem[Buldyrev et~al.(2010)Buldyrev, Parshani, Gerald, Stanley~H., and
  Havlin]{buldyrev:2010}
S.~V. Buldyrev, R.~Parshani, P.~Gerald, E.~Stanley~H., and S.~Havlin.
\newblock Catastrophic cascade of failures in interdependent networks.
\newblock \emph{Nature}, 464:\penalty0 1025--1028, 2010.

\bibitem[Gao et~al.(2011)Gao, Buldyrev, Havlin, and Stanley]{gao:2011}
J.~Gao, S.~V. Buldyrev, S.~Havlin, and H.~E. Stanley.
\newblock {{R}obustness of a network of networks}.
\newblock \emph{Phys. Rev. Lett.}, 107:\penalty0 195701, 2011.

\bibitem[Um et~al.(2011)Um, Minnhagen, and Kim]{um:2011}
J.~Um, P.~Minnhagen, and B.~J. Kim.
\newblock {{S}ynchronization in interdependent networks}.
\newblock \emph{Chaos}, 21:\penalty0 025106, 2011.

\bibitem[Aguirre et~al.(2014)Aguirre, Sevilla-Escoboza, Guti\'errez, Papo, and
  Buld\'u]{aguirre:2014}
J.~Aguirre, R.~Sevilla-Escoboza, R.~Guti\'errez, D.~Papo, and J.~M. Buld\'u.
\newblock {{S}ynchronization of interconnected networks: the role of connector
  nodes}.
\newblock \emph{Phys. Rev. Lett.}, 112:\penalty0 248701, 2014.

\bibitem[G{\'o}mez-Garde{\~n}es et~al.(2012)G{\'o}mez-Garde{\~n}es, Reinares,
  Arenas, and Floria]{gomezgardenes:2012}
J.~G{\'o}mez-Garde{\~n}es, I.~Reinares, A.~Arenas, and L.~M. Floria.
\newblock {{E}volution of cooperation in multiplex networks}.
\newblock \emph{Sci Rep}, 2:\penalty0 620, 2012.

\bibitem[Wang et~al.(2012)Wang, Szolnoki, and Perc]{wang:2012}
Z.~Wang, A.~Szolnoki, and M.~Perc.
\newblock {Evolution of public cooperation on interdependent networks: The
  impact of biased utility functions}.
\newblock \emph{Europhys.\ Lett.}, 97:\penalty0 48001, 2012.

\bibitem[Dickison et~al.(2012)Dickison, Havlin, and Stanley]{dickison:2012}
M.~Dickison, S.~Havlin, and H.~E. Stanley.
\newblock Epidemics on interconnected networks.
\newblock \emph{Phys. Rev. E}, 85:\penalty0 066109, 2012.

\bibitem[Saumell-Mendiola et~al.(2012)Saumell-Mendiola, Serrano, and
  Bogu\~n\'a]{saumell:2012}
A.~Saumell-Mendiola, M.~A. Serrano, and M.~Bogu\~n\'a.
\newblock Epidemic spreading on interconnected networks.
\newblock \emph{Phys. Rev. E}, 86:\penalty0 026106, 2012.

\bibitem[Ya\ifmmode~\breve{g}\else \u{g}\fi{}an and Gligor(2012)]{yagan:2012}
O.~Ya\ifmmode~\breve{g}\else \u{g}\fi{}an and V.~Gligor.
\newblock Analysis of complex contagions in random multiplex networks.
\newblock \emph{Phys. Rev. E}, 86:\penalty0 036103, 2012.

\bibitem[O'Brien and Herschlag(1999)]{obrien:1999}
P.~J. O'Brien and D.~Herschlag.
\newblock Catalytic promiscuity and the evolution of new enzymatic activities.
\newblock \emph{Chem.\ Biol.}, 6:\penalty0 R91--R105, 1999.

\bibitem[Copley(2003)]{copley:2003}
S.~D. Copley.
\newblock Enzymes with extra talents: moonlighting functions and catalytic
  promiscuity.
\newblock \emph{Curr. Opin. Chem. Biol.}, 7:\penalty0 265--272, 2003.

\bibitem[Babtie et~al.(2010)Babtie, Tokuriki, and Hollfelder]{babtie:2010}
A.~Babtie, N.~Tokuriki, and F.~Hollfelder.
\newblock What makes an enzyme promiscuous?
\newblock \emph{Curr. Opin. Chem. Biol.}, 14:\penalty0 200--207, 2010.

\bibitem[Khersonsky and Tawfik(2010)]{khersonsky:2010}
O.~Khersonsky and D.~S. Tawfik.
\newblock Enzyme promiscuity: a mechanistic and evolutionary perspective.
\newblock \emph{Annu. Rev. Biochem.}, 79:\penalty0 471--505, 2010.

\bibitem[Copley(2015)]{copley:2015}
S.~D. Copley.
\newblock An evolutionary biochemist's perspective on promiscuity.
\newblock \emph{Trends Biochem. Sci.}, 40:\penalty0 72--78, 2015.

\bibitem[Copley(2017)]{copley:2017}
S.~D. Copley.
\newblock Shining a light on enzyme promiscuity.
\newblock \emph{Curr. Opin. Struc. Biol.}, 47, 2017.

\bibitem[Schmidt et~al.(2003)Schmidt, Mundorff, Dojka, Bermudez, Ness,
  Govindarajan, Babbitt, Minshull, and Gerlt]{schmidt:2003}
D.~M.~Z. Schmidt, E.~C. Mundorff, M.~Dojka, E.~Bermudez, J.~E. Ness,
  S.~Govindarajan, P.~C. Babbitt, J.~Minshull, and J.~A. Gerlt.
\newblock Evolutionary potential of ($\beta$/$\alpha$) 8-barrels: functional
  promiscuity produced by single substitutions in the enolase superfamily.
\newblock \emph{Biochemistry}, 42:\penalty0 8387--8393, 2003.

\bibitem[Aharoni et~al.(2005)Aharoni, Gaidukov, Khersonsky, Gould, Roodveldt,
  and Tawfik]{aharoni:2005}
A.~Aharoni, L.~Gaidukov, O.~Khersonsky, S.~M. Gould, C.~Roodveldt, and D.~S.
  Tawfik.
\newblock The'evolvability'of promiscuous protein functions.
\newblock \emph{Nat. Genet.}, 37:\penalty0 73, 2005.

\bibitem[Amitai et~al.(2007)Amitai, Gupta, and Tawfik]{amitai:2007}
G.~Amitai, R.~D. Gupta, and D.~S. Tawfik.
\newblock Latent evolutionary potentials under the neutral mutational drift of
  an enzyme.
\newblock \emph{HFSP J.}, 1:\penalty0 67, 2007.

\bibitem[Bloom et~al.(2007{\natexlab{b}})Bloom, Raval, and Wilke]{bloom:2007}
J.~D. Bloom, A.~Raval, and C.~O. Wilke.
\newblock Thermodynamics of neutral protein evolution.
\newblock \emph{Genetics}, 175:\penalty0 255--266, 2007{\natexlab{b}}.

\bibitem[Yang et~al.(2016)Yang, Hong, Baier, Jackson, and Tokuriki]{yang:2016}
G.~Yang, N.~Hong, F.~Baier, C.~J. Jackson, and N.~Tokuriki.
\newblock Conformational tinkering drives evolution of a promiscuous activity
  through indirect mutational effects.
\newblock \emph{Biochemistry}, 55:\penalty0 4583--4593, 2016.

\bibitem[Baier et~al.(2017)Baier, Hong, Yang, Pabis, Barrozo, Carr, Kamerlin,
  Jackson, and Tokuriki]{baier:2017}
F.~Baier, N.~Hong, G.~Yang, A.~Pabis, A.~Barrozo, P.~D. Carr, S.~C. Kamerlin,
  C.~J. Jackson, and N.~Tokuriki.
\newblock Cryptic genetic variation defines the adaptive evolutionary potential
  of enzymes.
\newblock \emph{bioRxiv}, page 232793, 2017.

\bibitem[Paaby and Rockman(2014)]{paaby:2014}
A.~B. Paaby and M.~V. Rockman.
\newblock Cryptic genetic variation: evolution's hidden substrate.
\newblock \emph{Nature Rev.\ Gen.}, 15:\penalty0 247--258, 2014.

\bibitem[Hughes(1994)]{hughes:1994}
A.~L. Hughes.
\newblock The evolution of functionally novel proteins after gene duplication.
\newblock \emph{Proc. R. Soc. Lond. B}, 256:\penalty0 119--124, 1994.

\bibitem[Kim et~al.(2010)Kim, Kershner, Novikov, Shoemaker, and
  Copley]{kim:2010}
J.~Kim, J.~P. Kershner, Y.~Novikov, R.~K. Shoemaker, and S.~D. Copley.
\newblock Three serendipitous pathways in {E.~coli} can bypass a block in
  pyridoxal-5'-phosphate synthesis.
\newblock \emph{Mol. Syst. Biol.}, 6:\penalty0 436, 2010.

\bibitem[Bratulic et~al.(2015)Bratulic, Gerber, and Wagner]{bratulic:2015}
S.~Bratulic, F.~Gerber, and A.~Wagner.
\newblock Mistranslation drives the evolution of robustness in tem-1
  $\beta$-lactamase.
\newblock \emph{Proc.\ Natl.\ Acad.\ Sci.\ USA}, 112:\penalty0 12758--12763,
  2015.

\bibitem[Whitehead et~al.(2008)Whitehead, Wilke, Vernazobres, and
  Bornberg-Bauer]{whitehead:2008}
D.~J. Whitehead, C.~O. Wilke, D.~Vernazobres, and E.~Bornberg-Bauer.
\newblock The look-ahead effect of phenotypic mutations.
\newblock \emph{Biol. Direct}, 3:\penalty0 18, 2008.

\bibitem[Drummond and Wilke(2009)]{drummond:2009}
D.~A. Drummond and C.~O. Wilke.
\newblock The evolutionary consequences of erroneous protein synthesis.
\newblock \emph{Nat. Rev. Genet.}, 10:\penalty0 715, 2009.

\bibitem[Yanagida et~al.(2015)Yanagida, Gispan, Kadouri, Rozen, Sharon, Barkai,
  and Tawfik]{yanagida:2015}
H.~Yanagida, A.~Gispan, N.~Kadouri, S.~Rozen, M.~Sharon, N.~Barkai, and D.~S.
  Tawfik.
\newblock The evolutionary potential of phenotypic mutations.
\newblock \emph{PLoS Genet.}, 11:\penalty0 e1005445, 2015.

\bibitem[Vaidya and Lehman(2009)]{vaidya:2009}
N.~Vaidya and N.~Lehman.
\newblock One {RNA} plays three roles to provide catalytic activity to a group
  {I} intron lacking an endogenous internal guide sequence.
\newblock \emph{Nucleic Acids Res.}, 37:\penalty0 3981--3989, 2009.

\bibitem[Talini et~al.(2011)Talini, Branciamore, and Gallori]{talini:2011}
G.~Talini, S.~Branciamore, and E.~Gallori.
\newblock Ribozymes: Flexible molecular devices at work.
\newblock \emph{Biochimie}, 93:\penalty0 1998--2005, 2011.

\bibitem[Lau and Unrau(2009)]{lau:2009}
M.~W. Lau and P.~J. Unrau.
\newblock A promiscuous ribozyme promotes nucleotide synthesis in addition to
  ribose chemistry.
\newblock \emph{Chem. Biol.}, 16:\penalty0 815--825, 2009.

\bibitem[Hayden et~al.(2011)Hayden, Ferrada, and Wagner]{hayden:2011}
E.~J. Hayden, E.~Ferrada, and A.~Wagner.
\newblock Cryptic genetic variation promotes rapid evolutionary adaptation in
  an rna enzyme.
\newblock \emph{Nature}, 474:\penalty0 92, 2011.

\bibitem[Wagner(2014)]{wagner:2014}
A.~Wagner.
\newblock Mutational robustness accelerates the origin of novel {RNA}
  phenotypes through phenotypic plasticity.
\newblock \emph{Biophys. J.}, 106:\penalty0 955--965, 2014.

\bibitem[Elowitz et~al.(2002)Elowitz, Levine, Siggia, and Swain]{elowitz:2002}
M.~B. Elowitz, A.~J. Levine, E.~D. Siggia, and P.~S. Swain.
\newblock Stochastic gene expression in a single cell.
\newblock \emph{Science}, 297:\penalty0 1183--1186, 2002.

\bibitem[Raser and O'Shea(2005)]{raser:2005}
J.~M. Raser and E.~K. O'Shea.
\newblock Noise in gene expression: origins, consequences, and control.
\newblock \emph{Science}, 309:\penalty0 2010--2013, 2005.

\bibitem[Maheshri and O'Shea(2007)]{maheshri:2007}
N.~Maheshri and E.~K. O'Shea.
\newblock Living with noisy genes: how cells function reliably with inherent
  variability in gene expression.
\newblock \emph{Annu. Rev. Bioph. Biom.}, 36, 2007.

\bibitem[Ackermann(2015)]{ackermann:2015}
M.~Ackermann.
\newblock A functional perspective on phenotypic heterogeneity in
  microorganisms.
\newblock \emph{Nat. Rev. Microbiol.}, 13:\penalty0 497, 2015.

\bibitem[van Boxtel et~al.(2017)van Boxtel, van Heerden, Nordholt, Schmidt, and
  Bruggeman]{vanboxtel:2017}
C.~van Boxtel, J.~H. van Heerden, N.~Nordholt, P.~Schmidt, and F.~J. Bruggeman.
\newblock Taking chances and making mistakes: non-genetic phenotypic
  heterogeneity and its consequences for surviving in dynamic environments.
\newblock \emph{J.\ R.\ Soc.\ Interface}, 14:\penalty0 20170141, 2017.

\bibitem[Herranz and Cohen(2010)]{herranz:2010}
H.~Herranz and S.~M. Cohen.
\newblock {MicroRNAs} and gene regulatory networks: managing the impact of
  noise in biological systems.
\newblock \emph{Gene. Dev.}, 24:\penalty0 1339--1344, 2010.

\bibitem[Lehner(2008)]{lehner:2008}
B.~Lehner.
\newblock Selection to minimise noise in living systems and its implications
  for the evolution of gene expression.
\newblock \emph{Mol. Syst. Biol.}, 4:\penalty0 170, 2008.

\bibitem[Little et~al.(2013)Little, Tikhonov, and Gregor]{little:2013}
S.~C. Little, M.~Tikhonov, and T.~Gregor.
\newblock Precise developmental gene expression arises from globally stochastic
  transcriptional activity.
\newblock \emph{Cell}, 154:\penalty0 789--800, 2013.

\bibitem[Eldar and Elowitz(2010)]{eldar:2010}
A.~Eldar and M.~B. Elowitz.
\newblock Functional roles for noise in genetic circuits.
\newblock \emph{Nature}, 467:\penalty0 167, 2010.

\bibitem[Veening et~al.(2008)Veening, Smits, and Kuipers]{veening:2008}
J.-W. Veening, W.~K. Smits, and O.~P. Kuipers.
\newblock Bistability, epigenetics, and bet-hedging in bacteria.
\newblock \emph{Annu. Rev. Microbiol.}, 62:\penalty0 193--210, 2008.

\bibitem[Norman et~al.(2015)Norman, Lord, Paulsson, and Losick]{norman:2015}
T.~M. Norman, N.~D. Lord, J.~Paulsson, and R.~Losick.
\newblock Stochastic switching of cell fate in microbes.
\newblock \emph{Annu. Rev. Microbiol.}, 69, 2015.

\bibitem[S{\"u}el et~al.(2006)S{\"u}el, Garc\'{\i}a-Ojalvo, Liberman, and
  Elowitz]{suel:2006}
G.~M. S{\"u}el, J.~Garc\'{\i}a-Ojalvo, L.~M. Liberman, and M.~B. Elowitz.
\newblock An excitable gene regulatory circuit induces transient cellular
  differentiation.
\newblock \emph{Nature}, 440:\penalty0 545, 2006.

\bibitem[Balaban et~al.(2004)Balaban, Merrin, Chait, Kowalik, and
  Leibler]{balaban:2004}
N.~Q. Balaban, J.~Merrin, R.~Chait, L.~Kowalik, and S.~Leibler.
\newblock Bacterial persistence as a phenotypic switch.
\newblock \emph{Science}, 305:\penalty0 1622--1625, 2004.

\bibitem[Beaumont et~al.(2009)Beaumont, Gallie, Kost, Ferguson, and
  Rainey]{beaumont:2009}
H.~J. Beaumont, J.~Gallie, C.~Kost, G.~C. Ferguson, and P.~B. Rainey.
\newblock Experimental evolution of bet hedging.
\newblock \emph{Nature}, 462:\penalty0 90, 2009.

\bibitem[West-Eberhard(2003)]{west-eberhard:2003}
M.~J. West-Eberhard.
\newblock \emph{Developmental plasticity and evolution}.
\newblock Oxford University Press, 2003.

\bibitem[Jeffery(1999)]{jeffery:1999}
C.~J. Jeffery.
\newblock Moonlighting proteins.
\newblock \emph{Trends Biochem. Sci.}, 24:\penalty0 8--11, 1999.

\bibitem[Copley(2012)]{copley:2012}
S.~D. Copley.
\newblock Moonlighting is mainstream: paradigm adjustment required.
\newblock \emph{Bioessays}, 34:\penalty0 578--588, 2012.

\bibitem[Piatigorsky(2007)]{piatigorsky:2007}
J.~Piatigorsky.
\newblock \emph{Gene sharing and evolution: the diversity of protein
  functions}.
\newblock Harvard University Press Cambridge MA:, 2007.

\bibitem[Kortmann and Narberhaus(2012)]{kortmann:2012}
J.~Kortmann and F.~Narberhaus.
\newblock Bacterial {RNA} thermometers: molecular zippers and switches.
\newblock \emph{Nat. Rev. Microbiol.}, 10:\penalty0 255, 2012.

\bibitem[Garcia-Martin et~al.(2016)Garcia-Martin, Dotu, Fernandez-Chamorro,
  Lozano, Ramajo, Martinez-Salas, and Clote]{garcia-martin:2016}
J.~A. Garcia-Martin, I.~Dotu, J.~Fernandez-Chamorro, G.~Lozano, J.~Ramajo,
  E.~Martinez-Salas, and P.~Clote.
\newblock {RNAiFold2T:} constraint programming design of {thermo-IRES}
  switches.
\newblock \emph{Bioinformatics}, 32:\penalty0 i360--i368, 2016.

\bibitem[Payne and Wagner(2014{\natexlab{b}})]{payne:2014c}
J.~L. Payne and A.~Wagner.
\newblock Latent phenotypes pervade gene regulatory circuits.
\newblock \emph{BMC Syst. Biol.}, 8:\penalty0 64, 2014{\natexlab{b}}.

\bibitem[Espinosa-Soto et~al.(2011)Espinosa-Soto, Martin, and
  Wagner]{espinosa-soto:2011}
C.~Espinosa-Soto, O.~C. Martin, and A.~Wagner.
\newblock Phenotypic plasticity can facilitate adaptive evolution in gene
  regulatory circuits.
\newblock \emph{BMC Evol. Biol.}, 11:\penalty0 5, 2011.

\bibitem[Jim{\'e}nez et~al.(2017)Jim{\'e}nez, Cotterell, Munteanu, and
  Sharpe]{jimenez:2017}
A.~Jim{\'e}nez, J.~Cotterell, A.~Munteanu, and J.~Sharpe.
\newblock A spectrum of modularity in multi-functional gene circuits.
\newblock \emph{Mol.\ Sys.\ Biol.}, 13:\penalty0 925, 2017.

\bibitem[Via and Lande(1985)]{via:1985}
S.~Via and R.~Lande.
\newblock Genotype-environment interaction and the evolution of phenotypic
  plasticity.
\newblock \emph{Evolution}, 39:\penalty0 505--522, 1985.

\bibitem[Scheiner(1993)]{scheiner:1993}
S.~M. Scheiner.
\newblock Genetics and evolution of phenotypic plasticity.
\newblock \emph{Annu. Rev. Ecol. Syst.}, 24:\penalty0 35--68, 1993.

\bibitem[Ancel(2000)]{ancel:2000b}
L.~W. Ancel.
\newblock Undermining the {Baldwin} expediting effect: does phenotypic
  plasticity accelerate evolution?
\newblock \emph{Theor. Popul. Biol.}, 58:\penalty0 307--319, 2000.

\bibitem[Price et~al.(2003)Price, Qvarnstr{\"o}m, and Irwin]{price:2003}
T.~D. Price, A.~Qvarnstr{\"o}m, and D.~E. Irwin.
\newblock The role of phenotypic plasticity in driving genetic evolution.
\newblock \emph{Proc.\ R.\ Soc.\ Lond.\ B}, 270:\penalty0 1433--1440, 2003.

\bibitem[Lande(2009)]{lande:2009}
R.~Lande.
\newblock Adaptation to an extraordinary environment by evolution of phenotypic
  plasticity and genetic assimilation.
\newblock \emph{J. Evol. Biol.}, 22:\penalty0 1435--1446, 2009.

\bibitem[Scheiner and Holt(2012)]{scheiner:2012}
S.~M. Scheiner and R.~D. Holt.
\newblock The genetics of phenotypic plasticity. {X. Variation} versus
  uncertainty.
\newblock \emph{Ecol. Evol.}, 2:\penalty0 751--767, 2012.

\bibitem[G{\'o}mez-Mestre and Jovani(2013)]{gomez-mestre:2013}
I.~G{\'o}mez-Mestre and R.~Jovani.
\newblock A heuristic model on the role of plasticity in adaptive evolution:
  plasticity increases adaptation, population viability and genetic variation.
\newblock \emph{Proc. R. Soc. B}, 280:\penalty0 20131869, 2013.

\bibitem[Draghi and Whitlock(2012)]{draghi:2012}
J.~A. Draghi and M.~C. Whitlock.
\newblock Phenotypic plasticity facilitates mutational variance, genetic
  variance, and evolvability along the major axis of environmental variation.
\newblock \emph{Evolution}, 66:\penalty0 2891--2902, 2012.

\bibitem[McCaskill(1990)]{mccaskill:1990}
J.~S. McCaskill.
\newblock The equilibrium partition function and base pair binding
  probabilities for rna secondary structure.
\newblock \emph{Biopolymers}, 29:\penalty0 1105--1119, 1990.

\bibitem[Koonin(2011)]{koonin:2011b}
E.~V. Koonin.
\newblock Are there laws of genome evolution?
\newblock \emph{PLoS Comput.\ Biol.}, 7:\penalty0 e1002173, 2011.

\bibitem[Iranzo et~al.(2014)Iranzo, G\'omez, L\'opez~de Saro, and
  Manrubia]{iranzo:2014}
J.~Iranzo, M.~G\'omez, F.~L\'opez~de Saro, and S.~C. Manrubia.
\newblock Large-scale genomic analysis suggests a neutral punctuated dynamics
  of transposable elements in bacterial genomes.
\newblock \emph{PLoS Comput.\ Biol.}, 10:\penalty0 e1003680, 2014.

\bibitem[Iranzo et~al.(2017)Iranzo, Cuesta, Manrubia, Katsnelson, and
  Koonin]{iranzo:2017}
J.~Iranzo, J.~Cuesta, S.~Manrubia, M.~Katsnelson, and E.~Koonin.
\newblock Disentangling the effects of selection and loss bias on gene
  dynamics.
\newblock \emph{Proc.\ Natl.\ Acad.\ Sci.\ USA}, 114:\penalty0 E5616, 2017.

\bibitem[Sch\"onknecht et~al.(2014)Sch\"onknecht, Weber, and
  Lercher]{schonknecht:2014}
G.~Sch\"onknecht, A.~Weber, and M.~Lercher.
\newblock Horizontal gene acquisitions by eukaryotes as drivers of adaptive
  evolution.
\newblock \emph{BioEssays}, 36:\penalty0 9, 2014.

\bibitem[Koonin(2015)]{koonin:2015}
E.~V. Koonin.
\newblock The turbulent network dynamics of microbial evolution and the
  statistical tree of life.
\newblock \emph{J.\ Mol.\ Evol.}, 80:\penalty0 244--250, 2015.

\bibitem[Iranzo et~al.(2016{\natexlab{b}})Iranzo, Krupovic, and
  Koonin]{iranzo:2016b}
J.~Iranzo, M.~Krupovic, and E.~V. Koonin.
\newblock The double-stranded {DNA} virosphere as a modular hierarchical
  network of gene sharing.
\newblock \emph{mBio}, 7:\penalty0 e00978, 2016{\natexlab{b}}.

\bibitem[Luc\'{\i}a-Sanz and Manrubia(2017)]{lucia-sanz:2017}
A.~Luc\'{\i}a-Sanz and S.~Manrubia.
\newblock Multipartite viruses: Adaptive trick or evolutionary treat?
\newblock \emph{npj Sys.\ Biol.\ App.}, 3:\penalty0 34, 2017.

\bibitem[Ueda et~al.(2017)Ueda, Takeuchi, and Kaneno]{ueda:2017}
M.~Ueda, N.~Takeuchi, and K.~Kaneno.
\newblock Stronger selection can slow down evolution driven by recombination on
  a smooth fitness landscape.
\newblock \emph{PLoS ONE}, 12:\penalty0 e0183120, 2017.

\bibitem[Peabody et~al.(2017)Peabody, Li, and Kao]{peabody:2017}
G.~L. Peabody, H.~Li, and K.~C. Kao.
\newblock Sexual recombination and increased mutation rate expedite evolution
  of {E}scherichia coli in varied fitness landscapes.
\newblock \emph{Nature Communications}, 8:\penalty0 2112, 2017.

\bibitem[Crameri et~al.(1998)Crameri, Raillard, Bermudez, and
  Stemmer]{crameri:1998}
A.~Crameri, S.~Raillard, E.~Bermudez, and W.~Stemmer.
\newblock {DNA} shuffling of a family of genes from diverse species accelerates
  directed evolution.
\newblock \emph{Nature}, 391:\penalty0 288--291, 1998.

\bibitem[Azevedo et~al.(2006)Azevedo, Lohaus, Srinivasan, Dang, and
  Burch]{azevedo:2006}
R.~B. Azevedo, R.~Lohaus, S.~Srinivasan, K.~K. Dang, and C.~L. Burch.
\newblock Sexual reproduction selects for robustness and negative epistasis in
  artificial gene networks.
\newblock \emph{Nature}, 440:\penalty0 87, 2006.

\bibitem[Benton(1993)]{benton:1993}
M.~J. Benton.
\newblock \emph{The Fossil Record 2}.
\newblock Chapman {\&} Hall, London, 1993.

\bibitem[Harland et~al.(1990)Harland, Armstrong, Cox, Craig, Smith, and
  Smith]{harland:1990}
W.~B. Harland, R.~L. Armstrong, A.~V. Cox, L.~E. Craig, A.~G. Smith, and D.~G.
  Smith.
\newblock \emph{A Geologic Time Scale 1989}.
\newblock Cambridge University Press, Cambridge, 1990.

\bibitem[Garcia-Martin et~al.(2018)Garcia-Martin, Catal\'an, Cuesta, and
  Manrubia]{garcia-martin:2018}
J.~A. Garcia-Martin, P.~Catal\'an, J.~A. Cuesta, and S.~Manrubia.
\newblock Phenotype size distributions in exact enumerations of genotype
  spaces.
\newblock \emph{Europhys.\ Lett.}, 2018.

\end{thebibliography}
\end{document}